\documentclass[twocolumn]{aastex631}%

\usepackage{xspace}

\begin{document}

\title{ASTRAFier: A Novel and Scalable Transformer-based Stellar Variability Classifier}

\author[0009-0001-8405-1504]{Paul~F.~X.~Gregory}
\affiliation{MIT Kavli Institute for Astrophysics \& Space Research, Massachusetts Institute of Technology, Cambridge, MA, USA}

\author[0000-0002-4371-3460]{Jeroen~Audenaert}
\affiliation{MIT Kavli Institute for Astrophysics \& Space Research, Massachusetts Institute of Technology, Cambridge, MA, USA}
\correspondingauthor{Jeroen Audenaert (jeroena@mit.edu), Paul~F.~X.~Gregory (paulg9@mit.edu)}

\author[0000-0002-3334-9984]{Mykyta Kliapets}
\affiliation{MIT Kavli Institute for Astrophysics \& Space Research, Massachusetts Institute of Technology, Cambridge, MA, USA}
\affiliation{Institute of Astronomy, KU Leuven, Celestijnenlaan 200D, bus 2401, 3001 Leuven, Belgium}

\author[0000-0002-5788-9280]{Daniel~Muthukrishna}
\affiliation{MIT Kavli Institute for Astrophysics \& Space Research, Massachusetts Institute of Technology, Cambridge, MA, USA}
\affiliation{AstroAI, Center for Astrophysics $\vert$ Harvard \& Smithsonian, 60 Garden Street, Cambridge, 02138, MA, USA}

\author[0000-0003-0842-2374]{Andrew Tkachenko}
\affiliation{Institute of Astronomy, KU Leuven, Celestijnenlaan 200D, bus 2401, 3001 Leuven, Belgium}

\author[0000-0002-7602-0046]{Marek Skarka}
\affiliation{Astronomical Institute of Czech Academy of Sciences, Fričova 298, 251 65 Ondřejov, Czech Republic}

\author[0000-0003-2400-6960]{Marc~Hon}
\affiliation{Department of Physics, National University of Singapore, 21 Lower Kent Ridge Road, Singapore, 119077}
\affiliation{MIT Kavli Institute for Astrophysics \& Space Research, Massachusetts Institute of Technology, Cambridge, MA, USA}

\author[0000-0003-2058-6662]{George~R.~Ricker}
\affiliation{MIT Kavli Institute for Astrophysics \& Space Research, Massachusetts Institute of Technology, Cambridge, MA, USA}




\def\kepler{\textit{Kepler}\xspace}

\begin{abstract}

Photometric missions such as \textit{Kepler} and TESS have generated millions of light curves covering almost the entire sky, offering unprecedented opportunities to study stellar variability and advance our understanding of the Universe. In this data-rich environment, machine learning has emerged as a powerful tool to efficiently and accurately process and classify light curves according to their type of stellar variability. In this work, we introduce ASTRAFier: a novel Transformer-based model for variability classification that integrates Bidirectional Long Short-Term Memory (BiLSTM) and Convolutional Neural Networks (CNNs). The model operates directly on time series without requiring feature engineering, creating an easy-to-maintain and efficient end-to-end classification framework. We train and validate our model using both \textit{Kepler} and TESS light curves and, respectively, achieve a classification accuracy of $94.26\%$ on \textit{Kepler} and $88.22\%$ on TESS. We demonstrate scalability by deploying our model on $\sim 2.8$ million TESS light curves from sectors 14, 15, and 26 (Kepler Field-of-View) delivered by MIT's Quick-look Pipeline (QLP) and release the resulting stellar variability catalog.


\end{abstract}



\section{Introduction} \label{sec:intro}

The temporal variability of stars can reveal their interior workings, evolutionary pathways and the presence of companion objects, while large-scale analyses can provide insights for population studies \citep[e.g.,][]{asteroseismology,Aerts2021,Kurtz2022}. Stellar variability and asteroseismology have been revolutionized with the advent of space missions such as \textit{Kepler}/K2 \citep{Borucki2010,Koch2010,Howell2014} and the Transiting Exoplanet Satellite Survey \citep[TESS,][]{Ricker2015}, delivering millions of uninterrupted high-quality light curves \citep[e.g.,][]{Huber2025}.

By now, TESS has observed nearly the entire sky in sectors of 27.4 days. The Full-Frame Images (FFIs) have 30-min, 10-min and 200-sec cadence for the primary (PM), first extended (EM1) and second extended (EM2) mission, respectively. The total observing baselines ranging from a few months to multiple years in the Continuous Viewing Zone, where the recently started third extended mission (EM3) also includes a number of 54 day sectors. The upcoming PLAnetary Transits and Oscillations of stars \citep[PLATO,][]{Rauer2024} mission will be launched in 2027 and continuously observe the same patch of sky for at least two years of time \citep{Nascimbeni2025, Jannsen2025}.

The sheer scale of the data necessitates efficient and effective automated analysis methods \citep[e.g.,][]{Audenaert2025}. The classification of stars according to their variability type is essential for building large samples of stars for detailed astrophysical analyses, identifying promising targets for follow-up observations, and informing future space missions \citep[e.g.,][who studied the PLATO field-of-view using TESS]{Eschen2024}. 

Variability catalogs for TESS have been constructed using statistical and visual methods for subsets of TESS observations \citep[e.g.,][]{Skarka2022,Fetherolf2023,Skarka2024,Kemp2025}. Additionally, dedicated classification methodologies relying on machine learning and statistical techniques have been created for identifying solar-like oscillators \citep[e.g.,][]{Hon2018a,Hon2018b,Hon2019,Nielsen2022,Hatt2023}, eclipsing binaries \citep[e.g.,][]{IJspeert2021,IJspeert2024a,IJspeert2024b}, short-period variables \citep[e.g.,][]{Olmschenk2024}, transients \citep[e.g.,][]{Roxburgh2025} and pulsators \citep[e.g., using both TESS and Gaia,][]{Hey2024}.

Machine learning has proven to be the most effective technique for performing large-scale automated classifications across a wide range of variability classes \citep[e.g.,][]{JamalBloom2020,Audenaert2021,Huijse2025, Audenaert2025}. Traditionally, supervised classification methodologies mostly relied on feature engineering techniques to characterize the properties of light curves, for example, with features derived from statistical moments, Lomb-Scargle periodogram \citep{lomb1976,scargle1982} and entropy \citep{Shannon1948}, such as those in \citet{Choi2025}. The features are then fed as input to, for example, random forests \citep{Breiman:fb}, gradient boosting machines \citep{friedman2001}, Gaussian mixture models or Convolutional Neural Networks (CNN) \citep[e.g.,][]{Debosscher2007,Sarro2009, Blomme2011,Richards2011,KimBailer-Jones2016,Armstrong2016,Hon2018a,Barbara2022,Cui2024}. In addition to supervised approaches, unsupervised settings have been increasingly explored to handle the growing volume of data; for instance, \citet{Audenaert2022} used entropy-based features in an unsupervised setting, while \cite{Ranaivomanana2025} and \cite{Huijse2025} utilized dimensionality reduction and deep representation learning via autoencoders, respectively, to discover and classify variable sources without prior labeling. \citet{Audenaert2021} combined multiple distinct models, each relying on different feature sets, into an ensemble classification model to achieve a higher performance.

Automated representation learning models \citep[see][for an overview]{Audenaert2025} have been used to learn the characteristic features of light curves for variability classification. \citet{Naul2018, becker2025} used Recurrent Neural Networks (RNNs) to classify sparse light curves, while \citet{Muthukrishna2019} used RNNs with gated recurrent units (GRUs) to classify transients.

Since their introduction, Transformers \citep[][]{Vaswani2017} have become a cornerstone in Generative Artificial Intelligence (AI) and natural language processing, powering models such as ChatGPT \citep{GPT} and BERT \citep[][]{BERT}. Their success in NLP has spurred interest in applying Transformer architectures to time series data, where their capacity to learn dependencies and correlations between sequence elements offers promising advantages \citep{Qingsong}. \cite{Pan2024} used a transformer along with a CNN to predict log g values from light curves. The use of the transformer was found to increase performance of the model over just a CNN, especially in capturing long term dependencies, with other examples being \citet{Donoso-Oliva2023,Rizhko2025,Moreno-Cartagena2025,Donoso-Oliva2026}.

In this work, we present a novel machine learning framework to classify stars according to their variability classes. Our model, named ASTRAFier (Astronomical Sequence TRansformer-based vAriability classifier), utilizes LSTM, Transformers, and CNNs to process the light curve, offering a powerful architecture for classification. This architecture is designed to directly process raw light curve data, eliminating the need for feature engineering while effectively capturing the complex temporal patterns inherent in stellar variability. We build on the earlier classification work by the TESS Asteroseismic Science Consortium \citep{Audenaert2021} and leverage their training set of \textit{Kepler} light curves and its cross-match with TESS.

We give a theoretical overview of the different machine learning components in Sect.~\ref{sec:background}, discuss our model in Sect.~\ref{sec:architecture}, training set in Sect.~\ref{sect:trainingdata} and training procedure in Sect.~\ref{sec:training}. We analyze the results on our labeled training set in Sect.~\ref{sec:results} and deploy our model to all light curves in TESS sectors 14, 15, and 26 in Sect.~\ref{sec:deployment} to obtain a catalog of variable star candidates. These sectors are of particular interest as they provide spatial overlap with the Kepler field of view.


\section{Background}
\label{sec:background}
This section introduces the fundamental machine learning components behind our model in a light curve processing context: Transformers, CNNs, and LSTMs.

\subsection{Transformers}
\label{subsec:transformer}

The main mechanism behind the Transformer is multi-head self-attention \citep[MHSA,][]{Vaswani2017}. Self-attention works by transforming each token (a unit of data, for language models typically a word or part of a word, and in the case of light curves a time step) into three learned representations via linear projections; the query, key, and value. In short, the query seeks relevant context from other tokens. The key indicates how suitable a token is in responding to queries from other tokens. The value contains the content of the token that is weighted and aggregated based on how well the key matches the query, determining which parts of the original sequence influence the output. The attention matrix is computed as follows: 

\begin{equation}
\text{Attention} (\textbf{Q},\textbf{K},\textbf{V}) = \text{softmax}(\frac{\textbf{Q}\textbf{K}^{\text{T}}}{\sqrt{\text{d}_{\text{k}}}})\textbf{V},
\end{equation}
where \textbf{Q}, \textbf{K}, and \textbf{V} are the query, key, and value matrices, respectively, and $\text{d}_{\text{k}}$ is the dimension of the key matrix.

This computes the relevance of each position in the sequence to every other position, telling the model where it should pay more attention (i.e., the ``attention'' mechanism). For multi-head attention, multiple self-attention mechanisms are employed in parallel with independently learned key, query, and value matrices, and these outputs are then concatenated, enabling the model to learn more complex relationships as different heads can focus on different parts of the input. The parallelism of the multiple heads also allows for more efficient computations.

To incorporate the sequential order of a time series into the model, as Transformers are inherently permutation-invariant, positional encodings are added to the input embeddings. The original Transformer architecture \citep[][]{Vaswani2017} introduced sinusoidal positional encodings, which alternate sine and cosine functions of varying frequencies across adjacent dimensions to encode absolute position. This approach has two key advantages: it allows the model to extrapolate to sequence lengths longer than those seen during training, and the sinusoidal structure enables the model to learn relative positions through linear projections. 

However, the standard positional encoding assumes uniformly spaced inputs, which is not guaranteed for astronomical time series. Light curves often contain gaps due to spacecraft operations, data quality cuts, or observing constraints. To address this, we derive our positional encoding directly from the time vector of the input light curve rather than using integer position indices \citep[][]{ZuoTime}, ensuring that the encoding reflects the true temporal spacing between observations. We additionally scale the input to the sine and cosine functions in the positional encoding by $d_{\text{emb}}/T$, following \citet{FoumaniTime}, which prevents the positional encodings from becoming indistinguishable when the embedding dimension is small relative to the sequence length. We apply these two modifications to the original encoding of \citet{Vaswani2017}, yielding the following:

\begin{equation}
\label{equation:pe1}
    \mathbf{PE}_{(\text{pos}, 2i)} = \sin\!\left(\frac{\text{pos}}{10000^{\,2i/d_{\text{emb}}}} \cdot \frac{d_{\text{emb}}}{T}\right)
\end{equation}
\begin{equation}
\label{equation:pe2}
    \mathbf{PE}_{(\text{pos}, 2i+1)} = \cos\!\left(\frac{\text{pos}}{10000^{\,2i/d_{\text{emb}}}} \cdot \frac{d_{\text{emb}}}{T}\right)
\end{equation}
             
where $\text{pos}$ is the observation timestamp, $i \in \{0, \ldots, d_{\text{emb}}/2 - 1\}$ along the embedding dimension, $T$ is the number of time steps, and $d_{\text{emb}}$ is the embedding dimension. $\mathbf{PE}$ has shape $(T, d_{\text{emb}})$ and is added element-wise to the Transformer input, ensuring that temporal order information is preserved.

\begin{figure}[h!]
    \centering
    \includegraphics[width=0.6\linewidth]{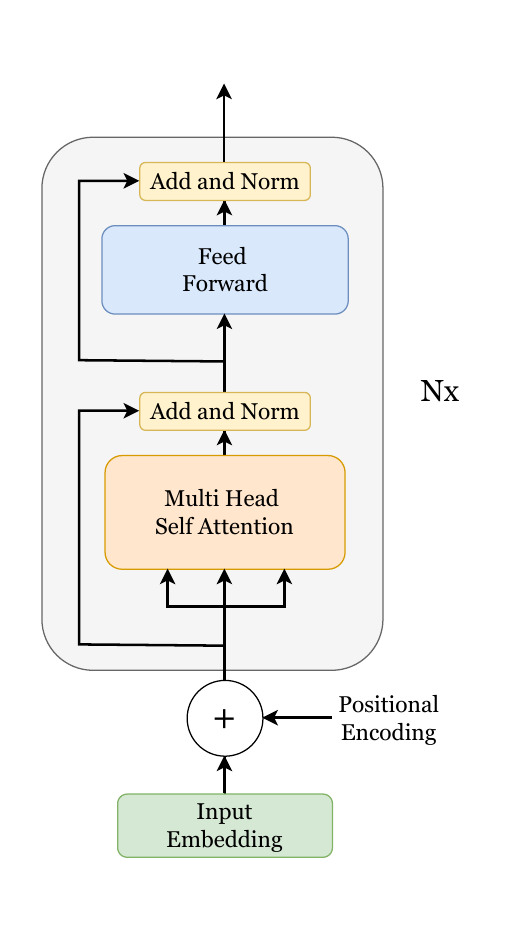}
    \caption{A Transformer encoder layer. Figure reproduced from \cite{Vaswani2017}.}
    \label{fig:transformer}
\end{figure}

A Transformer encoder block consists of a positional encoding followed sequentially by MHSA and a feed-forward (a network of non-linear transformations flowing in one direction) module. Residual connections are applied around both the self-attention and feed-forward modules, as illustrated in Fig.~\ref{fig:transformer}.

\subsection{Long Short-Term Memory (LSTM)} \label{subsec:lstm}
The foundation for LSTMs \citep{Hochreiter1997} was laid by Recurrent Neural Networks (RNNs). Unlike feedforward neural networks, RNNs are designed to process sequences of data by maintaining a hidden state that evolves over time. At each time step $t$, the network updates its hidden state $h_t$ based on the current input and the previous hidden state $h_{t-1}$. This recurrent connection allows the network to retain information from earlier time steps. A significant limitation of traditional RNNs is the vanishing gradient problem, where the influence of earlier inputs diminishes as gradients are backpropagated through many time steps, hindering the ability of RNNs to process long sequences.

\begin{figure}[h!]
    \centering
    \includegraphics[width=1\linewidth]{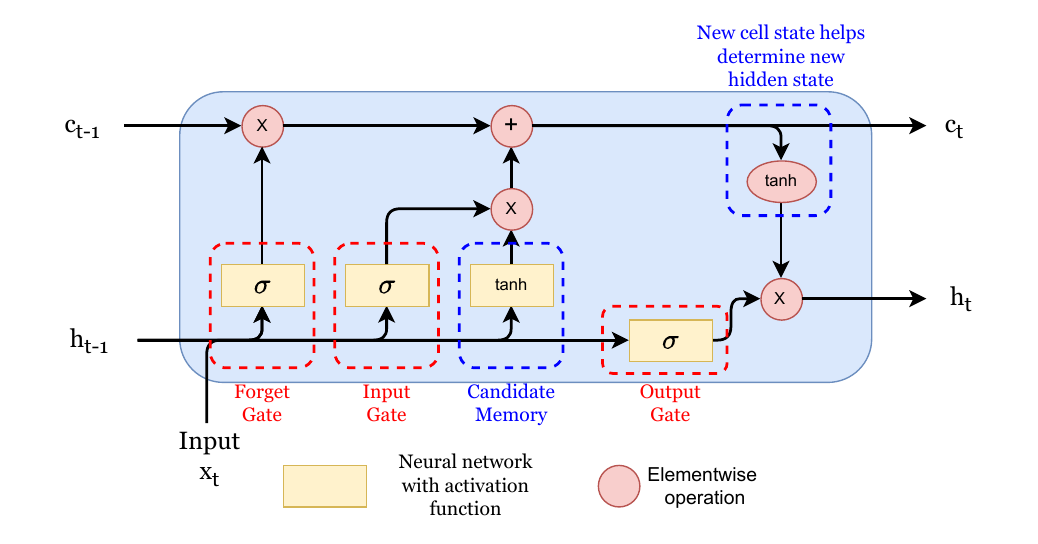}
    \caption{An LSTM block at time step t. $x_t$ is index t of the input sequence, $c_t$ is the cell state at time t, and $h_t$ is the hidden state at time t. An LSTM module consists of many such blocks, typically one for each time step in the input sequence. The LSTM outputs its hidden states $[h_1, h_2, ..., h_T]$.}
    \label{fig:LSTM}
\end{figure}

LSTMs address this issue through the use of a cell state that can retain important information over long durations, ensuring that long-term dependencies are not forgotten as the sequence progresses. In short, the cell state handles long-term memory, while the hidden state handles short-term memory. The LSTM uses three gates to control what information is remembered: the forget gate, the input gate, and the output gate. The forget gate determines what parts of the previous cell state can be discarded, the input gate decides how much of the new input should be added to the cell state, and the output gate regulates the influence of the cell state on the current hidden state. This gating mechanism enables LSTMs to preserve important information over extended sequences. An LSTM block for a single time step can be seen in Fig.~\ref{fig:LSTM}.

In our model, we make use of a bidirectional LSTM \citep[BiLSTM,][]{Schuster1997}. This expands on the LSTM by processing the input sequence in both the forward and backward directions. Essentially, one LSTM reads the sequence from start to end, another LSTM reads it from end to start, and these outputs are concatenated, allowing the model to leverage information from both past and future contexts. This dual perspective is particularly advantageous for time series classification, as it enables the capture of dependencies in both temporal directions.

\subsection{Convolutional Neural Networks (CNNs)} \label{subsec:cnn}
CNNs \citep{LeCun1998} are a feed-forward architecture proficient at handling grid-like data structures. Originally popularized in computer vision for tasks such as handwritten digit recognition \citep{LeCun1989, LeCun1998} and large-scale image classification \citep{Krizhevsky2012}, CNNs have also demonstrated significant utility in processing time-series data by treating sequences as one-dimensional grids to capture temporal patterns \citep[][]{Wang2016}. CNNs consist of convolutional layers that employ learnable filters, called kernels, to capture spatial hierarchies and extract local features from the input. The kernel slides across the input, transforming it with the values it has learned to produce the output. To handle the edges of the data, the input is padded with values, often zeros, that allow the center of the kernel to reach the edges. A standard CNN layer consists of a convolution, batch normalization, and an activation function. 

\begin{figure}[h!]
    \centering
    \includegraphics[width=1\linewidth]{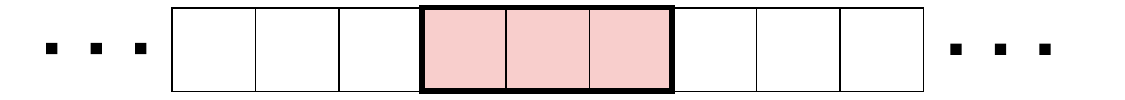}
    \caption{A 1-D kernel of size 3. This kernel slides along the input sequence \textit{stride} steps at a time, producing a new sequence through a multiplication of its learned weights and the input sequence.}
    \label{fig:cnn_ex}
\end{figure}

In time-series data, 1-D convolutions can be useful in detecting temporal patterns. An example of a 1-D kernel convolving an input sequence to produce a 1-channel output can be seen in Fig.~\ref{fig:cnn_ex}. 

While this example shows a CNN limited to handling 1-channel inputs and 1-channel outputs, we can generalize to handle multi-channel inputs and outputs as well. To handle a multi-channel input, we use a multi-channel kernel, filtering each channel in the input with the corresponding channel in the kernel, summing the outputs from each channel to get our 1-channel output. To handle a multi-channel output, we use a set of filters, called a filter bank. To get $C_{out}$ output channels, we use a filter bank of $C_{out}$ filters, one for each output channel. We combine these two ideas to be able to handle multi-channel inputs and outputs. 

\begin{figure}
    \centering
    \includegraphics[width=1\linewidth]{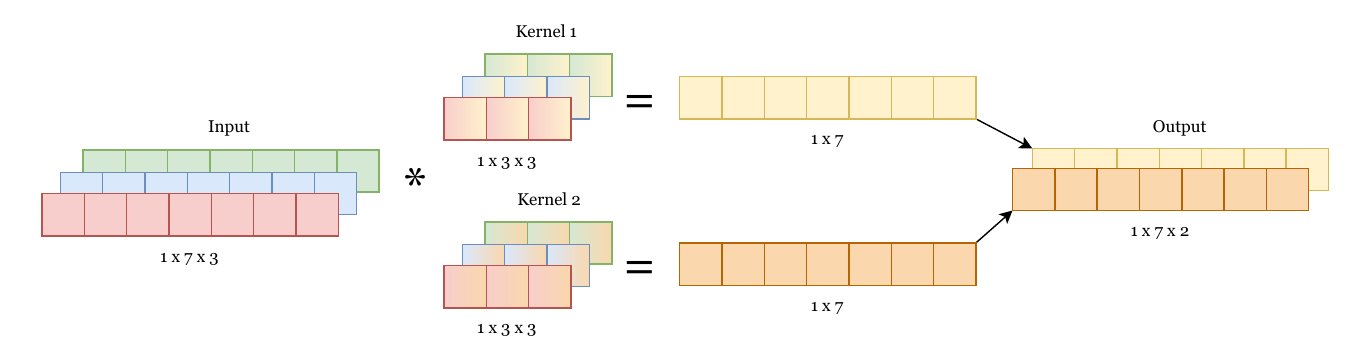}
    \caption{A visualization of a 1-D convolution with 3 input channels and 2 output channels.}
    \label{fig:multi_channel}
\end{figure}

As an example, take an input with $C_{in}$ channels $\textbf{x}_{in}\in \mathbb{R}^{C_{in} \times T}$. To get an output with $C_{out}$ channels, we compute \begin{equation}
    \textbf{x}_{out}[c_2, :] = \sum^{C_{in}}_{c_1=1}\textbf{w}[c_1,c_2,:]*\textbf{x}_{in}[c_1,:]
\end{equation} Where $\textbf{x}_{out} \in \mathbb{R}^{C_{out}\times T}$ is our output, $c_2\in \{0,...,C_{out}-1\} $ indexes the output channel, and $\textbf{w} \in \mathbb{R}^{C_{in}\times {C_{out}\times K}}$ is our filter bank of size \textit{K} kernels. Fig.~\ref{fig:multi_channel} shows a visualization of this.

In our CNN layers, we make use of the Gated Linear Unit activation function according to 

\begin{equation}
    GLU(a,b) = a \otimes \sigma(b)
\end{equation} 
where $a$ is the first half of the input and $b$ is the second half, and $\sigma$ is the sigmoid function. This activation function has been found to improve performance when modeling sequential data \citep[][]{GLU}.

\section{Model architecture}

\label{sec:architecture}
Recent research has shown the advantages of integrating attention mechanisms, CNNs, and LSTMs due to their complementary capabilities in handling sequential data \citep[e.g.,][]{Temperature,Stocks,Gasoline}. Transformers excel at capturing global context through self-attention, while CNNs specialize in detecting localized features via convolutional filters, and LSTMs manage sequential memory and long-term dependencies. 

\begin{figure}[h!]
    \centering
    \includegraphics[width=1\linewidth]{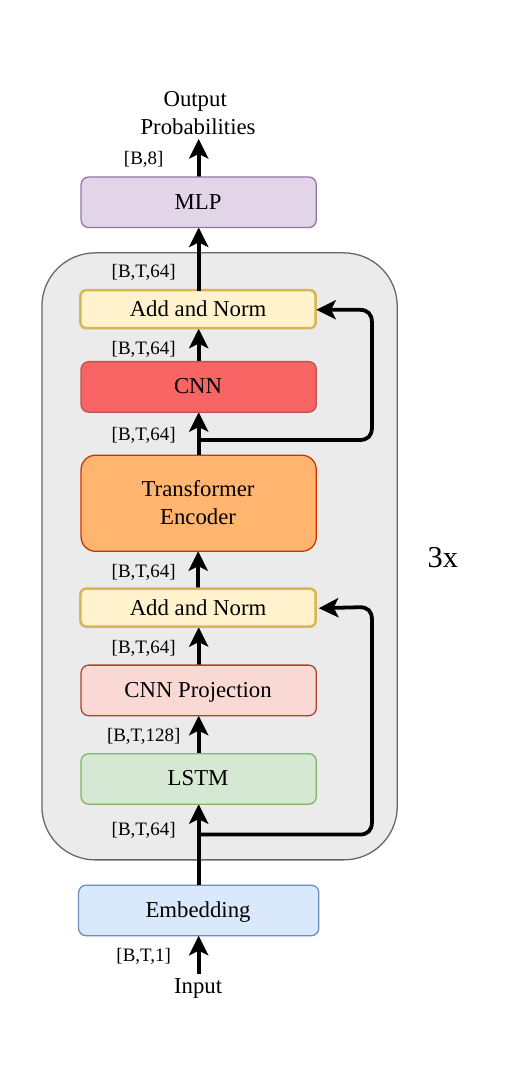}
    \caption{Architecture of the ASTRAFier model. The gray box highlights a single block, which is stacked three times. Each block contains a BiLSTM (with a CNN Projection), a Transformer encoder, and a CNN module. \textit{Add and Norm} refers to a residual (skip) connection followed by layer normalization, which facilitates gradient flow and stabilizes training. Positional information is injected into the initial embeddings using the time-dependent sinusoidal encodings described in Eqs.~\ref{equation:pe1}-\ref{equation:pe2}. Tensor shapes are annotated between modules in the form $[B, T, C]$, where $B$ is the batch size, $T$ the number of time steps, and $C$ the feature dimension.}
    \label{fig:architecture}
\end{figure}

Our model, ASTRAFier, is shown in Fig.~\ref{fig:architecture} and is a novel sequential hybrid architecture that integrates BiLSTM, Transformer, and CNN modules with residual connections. The residual connections add a module's input to its output and normalize the sum and are shown by the ``Add and Norm'' blocks. Our design enables each component to collaboratively process the information contained in a light curve, while the residual connections ensure that the characteristic information extracted by each module is preserved and effectively propagated throughout the network.

The light curve input is embedded and processed through three sequential ASTRAFier blocks (gray box in Fig.~\ref{fig:architecture}). The outputs of these blocks are then averaged across the time dimension and passed through a Multi-Layer Perceptron (MLP) with a softmax activation function for the final output probabilities for each class. 

\subsection{Handling Variable-Length Sequences}
\label{subsec:masking}
Astronomical light curves vary in length due to differences in observing strategy, instrument design and data quality flags. When processing batches of variable-length sequences, shorter sequences are padded to match the longest sequence in the batch. To prevent these padded positions from influencing the model's learned representations, we propagate binary padding masks throughout key stages of the network. These masks indicate valid observations ($m_t = 1$) versus padding ($m_t = 0$). Attention scores for padded positions are set to $-\infty$ before softmax, LSTM outputs at padded positions are zeroed, and final sequence representations are computed via masked averaging rather than standard global pooling. The convolutional layers do not explicitly apply the masks \citep[in line with, e.g.,][]{pytorchaudio}. Our normalization choice in these layers ensures the normalization for each valid frame is computed independently of any of the padded positions (see Sects.~\ref{subsec:lstmmodule} and \ref{subsec:cnn_module} for details). The residual effects are only confined to local boundary overlaps from the convolutional kernels, in contrast to a global effect that would be caused by unmasked attention. In the CNN Projection (Sect.~\ref{subsec:lstmmodule}), two of the three layers are pointwise and invariant to padding by construction, while the middle layer (kernel size 3) only sees zeros at boundary positions, since the BiLSTM output is already masked to zero at padded indices. In the CNN Module (Sect.~\ref{subsec:cnn_module}), the affected region scales with kernel size but remains local to frames near padding boundaries. The padded positions are excluded from the output entirely because the final representation uses masked averaging (Eq.~\ref{eq:masked_avg}).

\subsection{Embedding}
\label{subsec:embedding}

\begin{figure}[h!]
    \centering
    \includegraphics[width=1\linewidth]{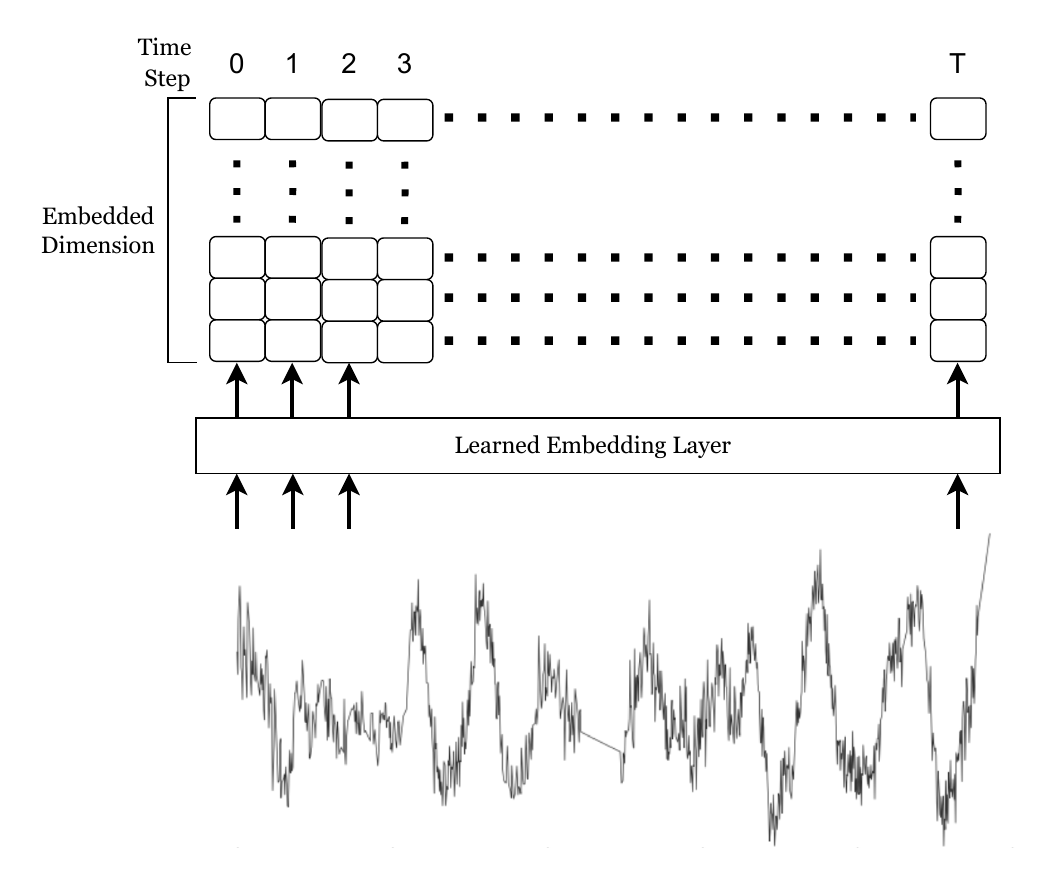}
    \caption{A preprocessed light curve is embedded into a higher dimension.}
    \label{fig:embedding}
\end{figure}

The light curves are first embedded using a fully connected layer that maps a single input feature (a scalar representing the flux at a particular time step) to 64 output features (our chosen embedding dimension). Essentially, this takes the input flux vector ($\mathbb{R}^{\mathrm{TimeSteps}}$), and for every time step, projects it into a 64-dimensional vector ($\mathbb{R}^{64}$) via a linear transformation, yielding a new embedded light curve ($\mathbb{R}^{TimeSteps \times 64}$). This higher dimensional representation allows the model to learn a richer representation of the data. Fig.~\ref{fig:embedding} illustrates the embedding process. 

\subsection{BiLSTM Module}
\label{subsec:lstmmodule}
Our embedded input is first passed through a 2-layer BiLSTM. Due to its bidirectional nature, the BiLSTM doubles the embedding dimension by concatenating features from both the forward and backward passes. After the BiLSTM layer, we apply the padding mask to suppress outputs corresponding to padded positions, preventing invalid time steps from influencing downstream computations. To reduce this expanded dimension back to our desired size while still retaining the bidirectional information, we employ a convolutional projection block, which we refer to as the CNN Projection in Fig.~\ref{fig:architecture} to distinguish it from the CNN Module described in Sect.~\ref{subsec:cnn_module}, as this block acts primarily as a channel reducer, projecting the $2 d_{\mathrm{emb}} = 128$ BiLSTM features down to $d_{\mathrm{emb}} = 64$. We found this to be more effective than simply halving the hidden dimension of the BiLSTM in each direction. Preliminary experiments demonstrated that the post BiLSTM convolutional projection approach yielded higher classification accuracy, likely because the learned kernels better preserve the salient features extracted by the BiLSTM passes. While this approach increases the computational complexity and parameter count, the gain in predictive performance justifies the additional overhead.

The convolutional projection block is composed of three sequential 1-D convolutions with Group Normalization \citep[\texttt{GroupNorm},][]{Wu_2018_ECCV} and ReLU activation: a pointwise convolution with 128 input channels and 256 output channels, a kernel size 3 convolution with 256 input channels and 256 output channels, and another pointwise convolution with 256 input channels and 64 output channels that returns the tensor to the set dimensionality. Two of the three layers are pointwise and act purely along the channel dimension, while the middle convolution (kernel size 3) additionally mixes information across neighboring time steps for further local temporal context. We use \verb"GroupNorm" rather than Batch Normalization \citep[\texttt{BatchNorm},][]{pmlr-v37-ioffe15} to improve stability with variable-length sequences and smaller batch sizes, as BatchNorm statistics can be unreliable when sequences contain varying amounts of padding. The output forms a residual connection with the input to the BiLSTM which is then normalized and passed into the Transformer. Beyond extracting useful features, the residual block serves as an additional form of positional encoding, enhancing the Transformer's ability to model temporal dependencies.

\subsection{Transformer Encoder}
In the Transformer layer, we use the positional encoding described in Sect.~\ref{subsec:transformer} and Eqs.~\ref{equation:pe1}-\ref{equation:pe2}, 8-head attention and replace the standard position-wise feed-forward network with a convolutional feed-forward module.
This substitution allows the feed-forward stage to incorporate local temporal context from neighboring time steps, rather than processing each position independently.

The attention mechanism incorporates the padding mask to ensure that attention is restricted to valid (non-padded) positions only. This is achieved by setting the attention scores for masked positions to negative infinity before the softmax operation, effectively zeroing their attention weights. The residual connection of the transformer encoder is then input to our CNN module.

\subsection{CNN Module}
\label{subsec:cnn_module}
\begin{figure}[h!]
    \centering
    \includegraphics[width=.8\linewidth]{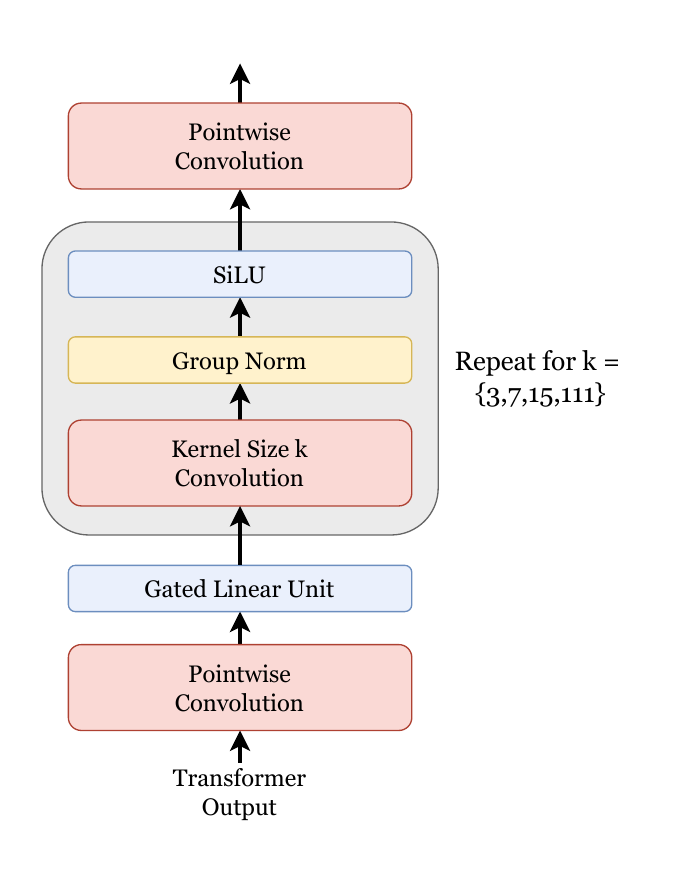}
    \caption{The CNN module.}
    \label{fig:cnn_module}
\end{figure}

The CNN module (Fig.~\ref{fig:cnn_module}) applies its convolutions along the time dimension, with the multi-scale kernels (sizes 3, 7, 15, and 111) capturing temporal patterns at progressively larger receptive fields. It passes the Transformer output through a pointwise convolution with 64 input channels and 256 output channels, followed by a Gated Linear Unit (GLU) which halves the channels to 128. This is followed by 4 convolutions of kernel sizes 3, 7, 15, and 111, each with 128 input and output channels and followed by a \verb"GroupNorm" and Sigmoid Linear Unit (SiLU) activation function. This range of different kernel sizes allows the model to capture both local and global trends from the Transformer's output. The specific combination of kernels was selected through an iterative optimization process on the validation set. As in the CNN Projection block, \verb"GroupNorm" (with a single group) is used throughout the module to maintain consistent normalization behavior regardless of the padding ratio within each batch. Lastly, another pointwise convolution is performed with 128 input channels and 64 output channels to return to the set dimensionality. The output forms a residual connection with the original CNN input, which is then normalized.

\subsection{Output Layer}
\label{subsec:output}
The final output probabilities are obtained by applying global average pooling across the time dimension. Because we work with variable-length sequences, we use masked averaging, that is, summing only over valid (non-padded) positions and dividing by the count of valid time steps. This ensures that padding tokens do not influence the learned representations. Formally, given the sequence output $\textbf{H} \in \mathbb{R}^{T\times d}$, where $T$ is the number of time steps and $d$ is the embedding dimension, and a binary mask $\textbf{m} \in \{0,1\}^T$ indicating the valid positions, the pooled representation is computed as:
\begin{equation}
    \mathbf{z} = \frac{\sum_{t=1}^{T} m_t \cdot \mathbf{h}_t}{\max\left(1, \sum_{t=1}^{T} m_t\right)}
    \label{eq:masked_avg}
\end{equation}

where $h_t$ is the $t$-th row of $\textbf{H}$. The pooled embedding is then passed through a 3-layer MLP for classification. The MLP consists of a linear layer projecting from 64 to 128 dimensions, followed by Layer Normalization \citep[\texttt{LayerNorm},][]{ba2016layernormalization}, SiLU activation, and dropout (p=0.2). A second linear layer reduces the dimension from 128 to 32, followed by SiLU activation and dropout (p=0.2). Finally, a linear layer maps from 32 dimensions to 8 class logits and is followed by a softmax function to produce output probabilities.

When performing inference, Monte-Carlo dropout \citep{MCD} is applied to calibrate probabilities and estimate predictive uncertainty. This consists of keeping dropout active during inference with a probability of 0.2 and performing 20 forward passes, taking the mean of these outputs as the final predicted probability distribution.

\section{Training data}
\label{sect:trainingdata}
Our training data consists of two labeled datasets: \textit{Kepler} light curves (Sect.~\ref{subsect:kepler_data}) and TESS QLP light curves (Sect.~\ref{subsubsec:dataselection}).

\subsection{Kepler}
\label{subsect:kepler_data}
We first validate the performance of our architecture on light curves from the \textit{Kepler} mission. We use the labeled benchmark dataset from \cite{Audenaert2021}, which consists of the following eight classes: (1) aperiodic variables (APERIODIC), (2) constant variables (CONSTANT), (3) contact binaries and rotational variables (CONTACT\_ROT), (4) $\delta$\,Scuti and $\beta$\,Cephei stars (DSCT\_BCEP), (5) eclipsing binaries and transit events (ECLIPSE), (6) $\gamma$\,Doradus and Slowly Pulsating B stars (GDOR\_SPB), (7) RR Lyrae and Cepheid variables (RRLYR\_CEPH), and (8) solar-like pulsators (SOLARLIKE). The detailed class descriptions are provided in \cite{Audenaert2021}. 

\subsection{TESS}
\label{subsubsec:dataselection}

We cross-match the \textit{Kepler} training set \citep{Audenaert2021} with TESS based on the TESS Input Catalog \citep{Stassun2018}. In order to increase the number of examples in challenging and smaller classes, we extend the RRLYR\_CEPH class with additional Cepheids and RR Lyraes from \citep{Ripepi2023Ceph,Clementini202RRLyr}, the DSCT\_BCEP, GDOR\_SPB and CONTACT\_ROT classes with the targets identified by \citet{Skarka2022,Skarka2024}. These samples were manually cleaned from ambiguous cases. We then retrieved the available MIT Quick-look Pipeline \citep[QLP,][]{QLP1,QLP2,QLP3,QLP4} light curves for the constructed catalog in sectors 14, 15 and 26 (\textit{Kepler FoV}), and perform visual inspections based on the light curves and periodograms and remove those light curves where no clear signal is found in the TESS QLP. Overall, there is a significant reduction in the number of unique targets because many of the light curves are dominated by noise and systematic properties that hide the astrophysical signatures. However, this is partially compensated by the inclusion of multiple sectors of data for the same target star. It is challenging to detect the oscillation and granulation patterns for solar-like oscillators based on single sector light curves, resulting in an overall reduction of the class size. Because of the large number of light curves dominated by systematic and instrumental trends, and in line with the findings from \citet{Audenaert2021,Tey2023}, we also add an INSTRUMENT/JUNK class to minimize confusion with astrophysical classes, and essentially replace the CONSTANT class with it because it consisted of simulated light curves. We populated the INSTRUMENT/JUNK class by selecting light curves from initial classification results that exhibited large recurring systematic trends or a lack of variability. The final training set is shown in Table~\ref{tab:trainingset}.

\begin{deluxetable*}{lcccccccccc}
\tabletypesize{\scriptsize}
\tablewidth{0pt} 
\tablecaption{The number of light curves for each class across our Kepler and TESS QLP datasets used in this work. \label{tab:train_pops}}
\tablehead{
\colhead{Dataset} & \colhead{APERIODIC} & \colhead{CONSTANT} & \colhead{CONTACT\_ROT} & \colhead{DSCT\_BCEP} & \colhead{ECLIPSE} & \colhead{GDOR\_SPB} & \colhead{INST./JUNK} & \colhead{RRLYR\_CEPH} & \colhead{SOLARLIKE} & \colhead{Total}}

\startdata 
{ Kepler } & 830 & 1000 & 2260 & 772 & 974 & 630 & 0 & 62 & 1800 & 8328\\ 
{ TESS QLP } & 1197 & 0 & 1489 & 1981 & 851 & 1085 & 1499 & 289 & 952 & 9343\\
\enddata
\end{deluxetable*}
\label{tab:trainingset}

\subsection{Light curve preprocessing}

We preprocess the light curves before feeding them to the model to remove noise and systematic trends in order to optimize performance, in line with \cite{Hey2024} and \cite{Kliapets2025}. We first remove the time steps and flux values flagged by QLP, remove NaN values and outlier values that deviate from the median by more than ten times the standard deviation. Subsequently, we run a 1-D Gaussian filter ($\sigma=61$) and subtract it from the light curve. This is because a Gaussian filter with high sigma value represents long range trends that are often present in TESS light curves but irrelevant to the stellar variability pattern. With this filter, the longest period that can pass through the Gaussian filter is 7.665 $d$ given the TESS sampling frequency $f_s = 48$ $d^{-1}$ ($0.02$ $d$) in the nominal mission. The relation between the standard deviation in time ($\sigma_t$) and frequency ($\sigma_f$) domains is then \begin{math}
    \sigma_f = \frac{0.02}{2\pi \sigma_t}
\end{math}. We note that the Gaussian filter could remove long-term stellar variability trends such as the year-long beating periods in g-mode pulsators previously found in Kepler by \citet{Beek2021}. Given those are not the primary focus of our research this is not an issue. Lastly, we shift the time values to start at zero in order to work better with the Transformer’s positional encoding and standardize the flux values. For \textit{Kepler} light curves, we apply only standardization, as the higher data quality requires less pre-processing compared to TESS.

\section{Training}
\label{sec:training}

We train our model using a batch size of 128 light curves. We use the AdamW optimizer \citep[][]{AdamW} with a learning rate ($\gamma$) of $10^{-4}$ for the ASTRAFier blocks and $10^{-3}$ for the MLP classification head, a weight decay coefficient ($\lambda$) of $1 \times 10^{-5}$, first and second moment decay rates ($\beta_1$ and $\beta_2$) of 0.9 and 0.95, respectively, and a numerical stability term ($\epsilon$) of $10^{-8}$. We use the AdamW optimizer due to its decoupled weight decay, which applies weight decay directly to the weights independent of gradient update. This leads to better convergence and regularization. To add further regularization, we use dropout layers \citep{Srivastava2014} with dropout probability of 0.2 throughout our model. We employ class-weighted cross-entropy loss with weights inversely proportional to class frequency ($w_c = N_{\mathrm{total}}/N_c$) to mitigate the effects of class imbalance. We split our data into $80\%$ training and $20\%$ holdout sets, stratified by class and split at the TIC level to ensure no target appears in both sets. During training, we further partition $10\%$ of the training set to obtain a validation set. We select the best-performing model based on validation accuracy and report its results on the holdout set.

\subsection{Computational complexity}
Our final model contains 8.8 million parameters with a size of 35 MB. We train on 2 × NVIDIA H200 GPUs, with each epoch completing in approximately 1.5 minutes for 12,038 training samples. On the two H200 GPUs, the classification of 1 million light curves with 20 forward passes each (for Monte Carlo dropout, see Sect.~\ref{subsec:output})  completes in approximately 8 hours.

\section{Results}
\label{sec:results}
We discuss the results of our model on the \textit{Kepler} and TESS datasets, validate our architectural choices and show the model's ability to accurately classify light curves.

We evaluate our architecture in three stages. We first train and test on \textit{Kepler} alone to benchmark against prior work (Sect.~\ref{subsec:kepler}), then train and test on TESS alone to assess performance with our refined class structure (Sect.~\ref{subsec:tess}), and finally train on combined \textit{Kepler} and TESS data but evaluating on the TESS holdout set, to produce our final deployment model (Sect.~\ref{subsec:tesskepler}).

\subsection{Kepler}
\label{subsec:kepler}

\begin{figure}
    \centering
    \includegraphics[width=\linewidth]{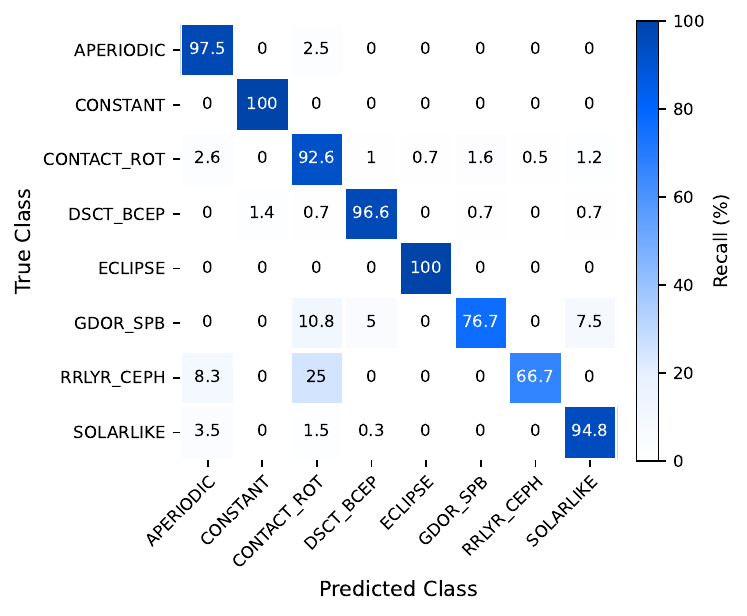}
    \caption{The confusion matrix on the \textit{Kepler} holdout set for the model trained only on \textit{Kepler} data.}
    \label{fig:kepler_conf}
\end{figure}

Training on our \textit{Kepler} dataset, we achieve a classification accuracy of $94.26\%$ on the holdout set. The confusion matrix for the holdout set is shown in Fig.~\ref{fig:kepler_conf}, and the recall, precision, and F1 scores\footnote{We calculate these metrics as follows: recall = $\frac{\text{TP}}{\text{TP+FN}}$, precision = $\frac{\text{TP}}{\text{TP+FP}}$, and F1 = $\frac{2*\text{recall}*\text{precision}}{\text{recall+precision}}$ where TP is the number of true positives for a class, FP the number of false positives, and FN the number of false negatives.}
for each class are shown in Table~\ref{tab:kepler_performance}. The final estimates are computed by averaging over the scores of the eight classes.

The overall accuracy is comparable to that of \cite{Audenaert2021}, who achieved an accuracy of $94.90\%$ on a different holdout set. Comparing the performance on a class-by-class basis, our model fails to correctly identify stars from the GDOR\_SPB class more often than in \cite{Audenaert2021}, with most of the confusion being with the CONTACT\_ROT class, a well-known challenge in variability classification (e.g., \citealt{Audenaert2021}; \citealt{Barbara2022}; \citealt{Hey2024}). Our model also fails to correctly identify the RRLYR\_CEPH class more often with most of the confusion being again with the CONTACT\_ROT class. However, there are only 12 RRLYR\_CEPH stars in our holdout set, making it less reliable. Our model is better able to identify eclipses, achieving a recall of $100\%$. The remaining classes are all within $1\%$ when comparing results. It should be noted again that the results from \citet{Audenaert2021} are on a different training and holdout set split, so comparisons should be interpreted with caution.

\begin{deluxetable}{lccccc}
\tabletypesize{\scriptsize}
\tablewidth{0pt} 
\tablecaption{Performance of the model trained on \textit{Kepler} data, on the Kepler holdout set (in \%). \label{tab:kepler_performance}}
\tablehead{
\colhead{Class} &  \colhead{Recall} & \colhead{Precision} & \colhead{F1}}
\startdata 
{   APERIODIC    } &   97.47(154/158) &  86.52(154/178) & 91.67 \\ 
{    CONSTANT   }  & 100.00(190/190) &  98.96(190/192) & 99.48\\
{   CONTACT\_ROT  }  &  92.56(398/430) &    93.87(398/424) & 93.21\\
{    DSCT\_BCEP   }  &    96.58(141/146)  &  92.76(141/152) & 94.63\\
{    ECLIPSE   }  & 100.00(185/185) &  98.40(185/188) & 99.20\\
{   GDOR\_SPB    }  &    76.67(92/120) &   92.00(92/100) & 83.64\\
{   RRLYR\_CEPH    }  &    66.67(8/12) &    80.00(8/10) & 72.73\\ 
{    SOLARLIKE   }  & 94.75(325/343) &  95.58(325/340) & 95.17\\
\hline
\hline
{    Total   }& 90.59  &  92.26 & 91.21\\
\hline
\hline
{Overall Accuracy} & \multicolumn{3}{c}{94.26} \\
\enddata
\end{deluxetable}

\subsection{TESS}
\label{subsec:tess}
\begin{figure}
    \centering
    \includegraphics[width=\linewidth]{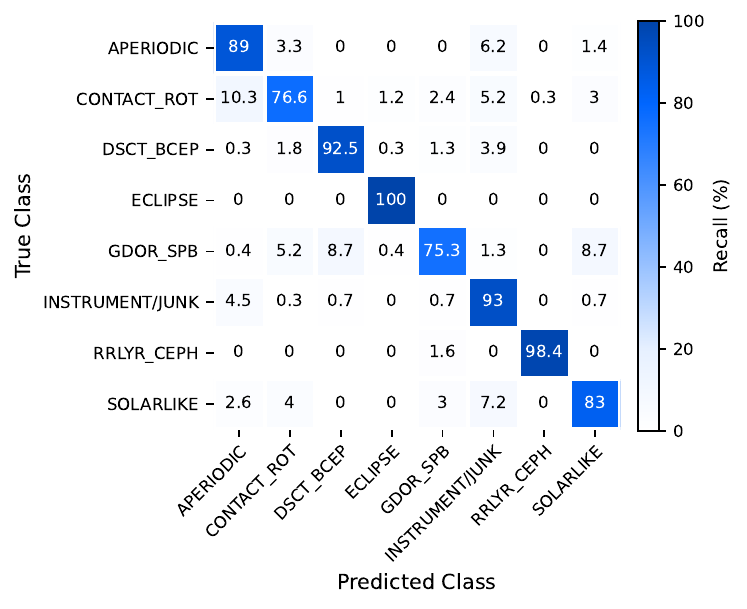}
    \caption{The confusion matrix on the TESS holdout set for the model trained only on TESS data.}
    \label{fig:cm_tess_only}
\end{figure}

Using the refined class structure described in Sect.~\ref{subsubsec:dataselection}, we first evaluate our model trained exclusively on the TESS dataset. On the TESS holdout set, we achieve a classification accuracy of 87.09\%. The confusion matrix is shown in Fig.~\ref{fig:cm_tess_only}, and per-class metrics are reported in Table~\ref{tab:tess_only_metrics}. The most significant source of confusion is between the CONTACT\_ROT and APERIODIC classes, where $10.3\%$ of CONTACT\_ROT stars are misclassified as APERIODIC. This is likely due to rotational variables whose periods exceed or are comparable to the 27.4-day sector baseline, resulting in light curves that lack clear periodicity and are thus difficult to distinguish from aperiodic variability.

The introduction of the INSTRUMENT/JUNK class proves effective, capturing instrumental and non-variable artifacts with $93.13\%$ accuracy while limiting contamination of the astrophysical classes. The DSCT\_BCEP and RRLYR\_CEPH classes also perform well, with accuracies of $92.49\%$ and $98.44\%$, respectively, although for RRLYR\_CEPH we only have 64 light curves across 49 targets.

We achieve $100\%$ accuracy on our ECLIPSE class; however, we do not expect this to hold in deployment. The holdout set contains only 153 light curves from 100 unique targets, and examining the prediction probabilities, 142 of the 153 eclipses in the holdout set receive a confidence above $95\%$ (median $95.71\%$), indicating that these are predominantly unambiguous eclipsing signals. We expect misclassifications to occur on the full dataset, where less obvious or noisier eclipses will present more challenging cases.

\begin{deluxetable}{lccccc}
\label{tab:tess_only_metrics}
\tabletypesize{\scriptsize}
\tablewidth{0pt} 
\tablecaption{Performance of the model trained on TESS data, on the TESS holdout set (in \%). \label{tab:tess_only_performance}}
\tablehead{
\colhead{Class}  & \colhead{Recall} & \colhead{Precision} & \colhead{F1}}
\startdata 
{   APERIODIC    }  &   89.10(188/211) &  77.69(188/242) & 83.00 \\ 
{   CONTACT\_ROT  }  &  76.60(252/329) &    87.80(252/287) & 81.82\\
{    DSCT\_BCEP   }  &    92.49(357/386)  &  93.46(357/382) & 92.97\\
{    ECLIPSE   }  & 100.00(153/153) &  96.23(153/159) & 98.08\\
{   GDOR\_SPB    }  &    75.32(174/231) &   88.78(174/196) & 81.50\\
{   INSTRUMENT/JUNK    }  &    93.13(271/291) &    81.38(271/333) & 86.86\\
{   RRLYR\_CEPH    }  &    98.44(63/64) &    98.44(63/64) & 98.44\\ 
{    SOLARLIKE   }  & 82.99(161/194) &  82.14(161/196) & 82.56\\
\hline
\hline
{    Total   }& 88.51  &  88.24 & 88.15\\
\hline
\hline
{Overall Accuracy} & \multicolumn{3}{c}{87.09} \\
\enddata
\end{deluxetable}

\subsection{TESS and \textit{Kepler}}
\label{subsec:tesskepler}
Using the refined class structure, we trained our final deployment model on a combined dataset of TESS (Sect.~\ref{subsubsec:dataselection}) and Kepler (Sect.~\ref{subsect:kepler_data}) light curves, removing any Kepler targets already present in the TESS set to avoid data leakage. Since our goal is deployment on TESS, we evaluate on the TESS holdout set.

\begin{figure}
    \centering
    \includegraphics[width=\linewidth]{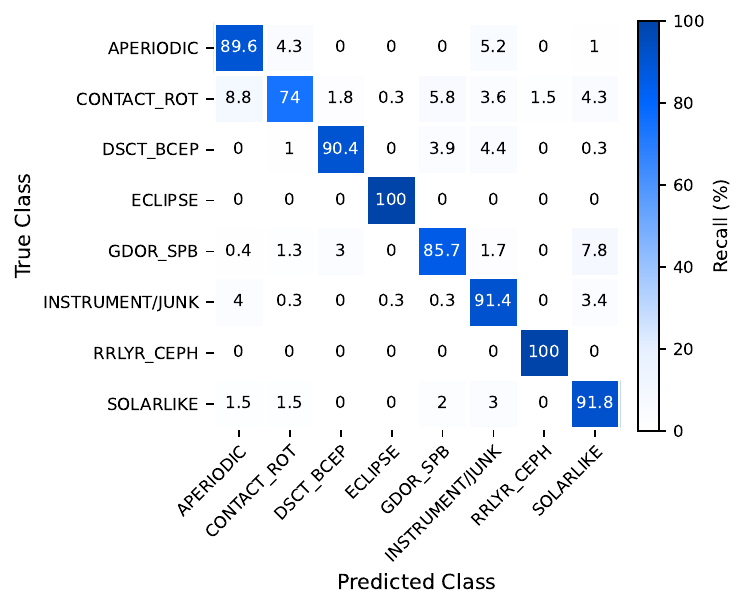}
    \caption{The confusion matrix on the TESS holdout set for the final model trained on both \textit{Kepler} and TESS data.
    }
    \label{fig:final_conf}
\end{figure}

\begin{figure}
    \centering
    \includegraphics[width=\linewidth]{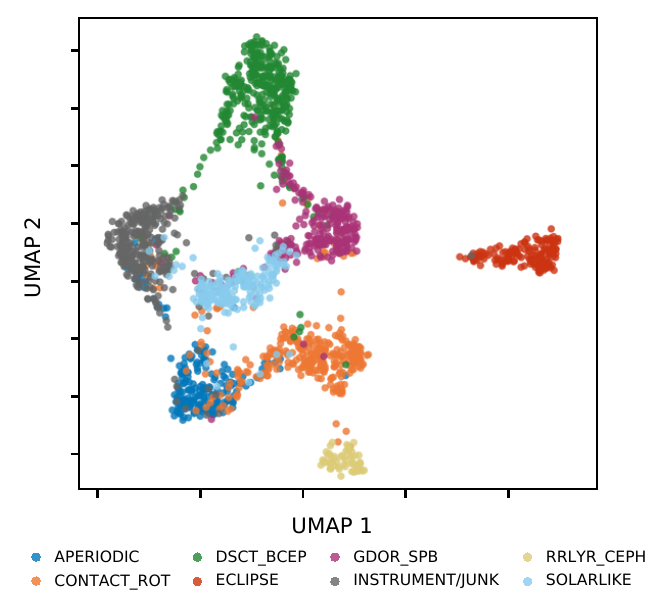}
    \caption{The UMAP reduction of the data points in our final QLP holdout set extracted before the final MLP layer.}
    \label{fig:umap}
\end{figure}

On the TESS holdout set, we achieve a classification accuracy of 88.22\%. The confusion matrix is shown in Fig.~\ref{fig:final_conf}, the UMAP visualization of holdout set embeddings in Fig.~\ref{fig:umap}, and per-class metrics in Table~\ref{tab:final_performance}. 

As in Sect.~\ref{subsec:tess}, the ECLIPSE class achieves perfect recall; we again attribute this to the holdout set containing predominantly unambiguous eclipsing signals rather than expecting this to generalize to deployment, and noting the limited holdout set size of 153 light curves across 100 targets. The RRLYR\_CEPH class also achieves $100\%$ recall, though with only 64 light curves from 49 unique targets in the holdout set, this should be interpreted with caution.

As shown in Table~\ref{tab:f1_tess_vs_combined}, we also demonstrate increased performance when adding \textit{Kepler} light curves to our training set, with F1 scores improving for seven of eight classes and a macro-averaged F1 gain of $+1.04$. The most notable improvements are in GDOR\_SPB ($+3.11$) and SOLARLIKE ($+2.81$). The only notable decrease is RRLYR\_CEPH ($-2.20$), which we attribute to the small class size making precision sensitive to even a few additional false positives. 
These gains demonstrate that our model scales effectively with training set size, suggesting further improvements are achievable as more labeled data becomes available. 

We additionally validate our architectural choices through an ablation study (Table~\ref{tab:ablation}), in which we remove one of the three core modules (BiLSTM, Attention, or CNN) from ASTRAFier and retrain; the full architecture outperforms all reduced variants, confirming that each component contributes meaningfully to classification performance.

\begin{deluxetable}{lccccc}
\tabletypesize{\scriptsize}
\tablewidth{0pt} 
\tablecaption{Performance of the final model trained on \textit{Kepler} and TESS data, on the TESS holdout set (in \%). \label{tab:final_performance}}
\tablehead{
\colhead{Class}  & \colhead{Recall} & \colhead{Precision} & \colhead{F1}}
\startdata 
{   APERIODIC    }  & 89.57(189/211) & 80.77(189/234) & 84.95\\ 
{   CONTACT\_ROT  }  & 73.86(243/329) & 92.40(243/263) & 82.10\\
{   DSCT\_BCEP    }  & 90.41(349/386) & 96.41(349/362) & 93.31\\ 
{   ECLIPSE      }  & 100.00(153/153) & 98.71(153/155) & 99.35\\
{   GDOR\_SPB     }  & 85.71(198/231) & 83.54(198/237) & 84.61\\
{   INSTRUMENT/JUNK   }  & 91.41(266/291) & 84.18(266/316) & 87.64\\ 
{   RRLYR\_CEPH    }  & 100.00(64/64) & 92.75(64/69) & 96.24\\
{   SOLARLIKE  }  & 91.75(178/194) & 79.82(178/223) & 85.37\\ 
\hline
\hline
{   Total        }& 90.34 & 88.57 & 89.20 \\
\hline
\hline
{Overall Accuracy} & \multicolumn{3}{c}{88.22} \\
\enddata
\end{deluxetable}

\begin{deluxetable}{lccc}
\tabletypesize{\scriptsize}
\tablewidth{0pt}
\tablecaption{Comparison on our TESS holdout set of F1 scores between the TESS-only model and the final model trained on combined TESS and \textit{Kepler} data.} \label{tab:f1_tess_vs_combined}
\tablehead{
\colhead{Class} & \colhead{TESS-only F1} & \colhead{TESS+\textit{Kepler} F1} & \colhead{Change}}
\startdata
{   APERIODIC    }& 83.00 & 84.95 & $+1.95$ \\
{   CONTACT\_ROT  }& 81.82 & 82.10 & $+0.28$ \\
{    DSCT\_BCEP   }& 92.97 & 93.31 & $+0.34$ \\
{    ECLIPSE   }& 98.08 & 99.35 & $+1.27$ \\
{   GDOR\_SPB    }& 81.50 & 84.61 & $+3.11$ \\
{   INSTRUMENT/JUNK    }& 86.86 & 87.64 & $+0.78$ \\
{   RRLYR\_CEPH    }& 98.44 & 96.24 & $-2.20$ \\
{    SOLARLIKE   }& 82.56 & 85.37 & $+2.81$ \\
\hline
\hline
{   Macro F1   }& 88.15 & 89.20 & $+1.04$ \\
\enddata
\end{deluxetable}

\begin{deluxetable}{lcccc}
\tabletypesize{\scriptsize}
\tablewidth{0pt}
\tablecaption{Ablation study on the combined TESS and Kepler holdout set. Each row removes one core module from ASTRAFier. \label{tab:ablation}}
\tablehead{
\colhead{Model} & \colhead{Accuracy} & \colhead{Recall} & \colhead{Precision} & \colhead{F1}}
\startdata
{ ASTRAFier }& 88.22 & 90.34 & 88.57 & 89.20 \\
{ Attention + CNN }& 87.14 & 88.76 & 86.73 & 87.44\\
{ BiLSTM + Attention }& 86.66 & 88.24 & 87.30 & 87.66\\
{ BiLSTM + CNN }& 85.48 & 87.37 & 85.84 & 86.35\\
\enddata
\tablecomments{For this experiment, we remove one module (LSTM, Attention, CNN) from our ASTRAFier model, keeping everything else the same. When we remove LSTM, we also remove the 3-layer CNN in its residual block. Recall, precision, and F1 are the macro-averaged score across all classes.}
\end{deluxetable}

We can visualize how well the model separates different classes by extracting the embeddings from before the final MLP layers and plotting them in 2 dimensions using UMAP \citep[][]{umap}. Examining the plot, there is noticeable overlap between the DSCT\_BCEP and GDOR\_SPB classes. These could be hybrid pulsating stars that exhibit both p and g modes \citep[e.g.,][]{fritzewski2025b,Kliapets2025}. We can also see confusion between the CONTACT\_ROT and APERIODIC classes, as well as the GDOR\_SPB and SOLARLIKE classes; which is consistent with the confusion matrix in Fig.~\ref{fig:final_conf}.

\section{Deploying the Classifier}
\label{sec:deployment}

We deploy our final model trained on \textit{Kepler} and TESS data on all $\sim 2.8$ million QLP light curves observed in TESS sectors 14, 15 and 26. In general, we find that the accuracy scores are in line with the reported testing scores in Sect.~\ref{sec:results}. However, because a training set is never a perfect representation of reality (e.g., small class sizes, varying systematics,...), there are always differences between testing and deployment performance on the full dataset.

We therefore evaluate the performance of our variability classification architecture during deployment by analyzing the astrophysical properties exhibited by sub-populations assigned a certain class, and support this with detailed inspections of their light curves and amplitude spectra.

\begin{figure*}
    \centering
    \includegraphics[width=18cm]{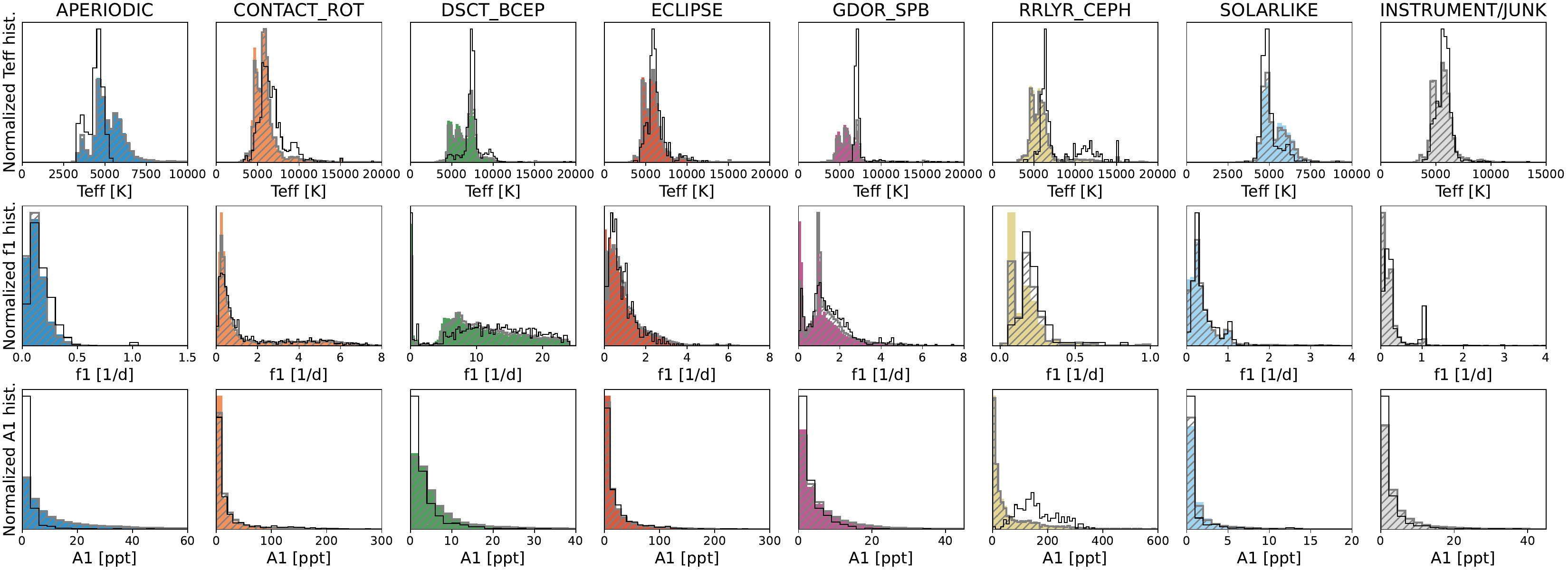}
    \caption{Normalized distributions of effective temperatures (top), dominant variability (middle), and its amplitude (bottom) of the labeled set (black outline) and classified targets from TESS Sectors 14, 15, and 26 with probabilities higher than 0.5 (color) and 0.8 (hash). If a star had more than one light curve in the \textit{Kepler} field of view, only the one with the higher probability was plotted. Distributions have been clipped on the right for visibility at different values for each class.}
    \label{fig:populations}
\end{figure*}

In Fig. \ref{fig:populations}, we show the distributions of effective temperature ($T_{\mathrm{eff}}$ from \textit{Gaia} DR3 data, top row), dominant variability $f_1$ (middle row), and its amplitude $A_1$ (bottom row) — on which the classifier had no prior information. We only plotted targets that received final scores above 0.5 per class; we revisit this later in this Section. We note that the frequencies and amplitudes of the OBAF-type pulsators are mostly in line with those in \cite{Hey2024} and \cite{aerts2025}. One notable exception is the distribution of $A_1$ for RRLYR\_CEPH class, which for the unlabeled TESS data is shifted much further to the left and peaks at very low amplitudes despite high performance demonstrated in Sect.~\ref{subsec:tesskepler}. This could be explained by this class being the least represented in the training data, leading to issues in successfully generalizing to unseen data while differentiating this class from rotational variables. This was supported by our manual inspection of random light curves that received high probabilities for this class. The rotation periods (inverses of $f_1$) for the CONTACT\_ROT class are comparable to the ones reported by \cite{colman2024} and are biased towards short periods, as expected from the TESS data. Other potential misclassifications revealed by the distributions are the confusion of solar-like oscillators with g-mode pulsators where $T_{\mathrm{eff}} < 6500$ K, as well as stars in the first peak of the bimodal $A_1$ distribution — typically slower-rotating g-mode pulsators — and which is higher in the unlabeled data than the labeled set, which coincides with the $A_1$ distribution peak for solar-like oscillators, consistent with Fig. \ref{fig:final_conf}.

\begin{figure}
    \centering
    \includegraphics[width=\linewidth]{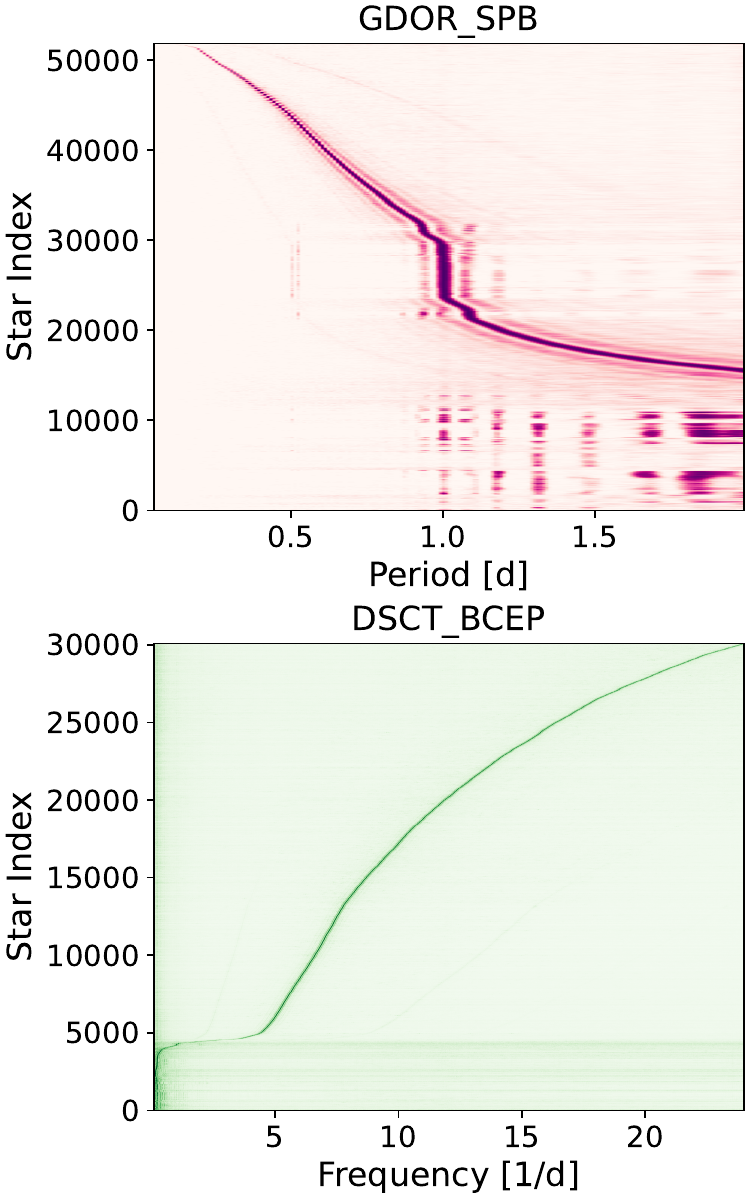}
    \caption{Stacked amplitude spectra of candidate g-mode (top, in period) and p-mode pulsators (bottom, in frequency), for which the prediction probability is higher than 0.5. Stars of each of the two classes are sorted by the dominant variability.}
    \label{fig:stacked}
\end{figure}

The latter is further supported by the analysis of amplitude spectra. In Fig. \ref{fig:stacked}, we show the stacked periodograms for stars labelled as g- or p-mode pulsators by the classifier. For g-mode pulsators, similar to \cite{li2020} and \cite{Hey2024}, we see a main ridge on the stacked amplitude spectra (top plot, in period), mostly populated by the prograde dipole ($(l,m) = (1,1)$) mode \citep{aerts2024}. The secondary lower ridge is likely associated with a lower-amplitude with $l=2$ or a harmonic of a dominant mode \citep{Hey2024}. Some targets also show potential r modes similar to those from \cite{li2020}. Stars immediately below the main ridge are once again likely misclassified solar-like oscillators. On the bottom of the plot, we see stars with clear harmonic behaviors, likely rotational variables or eclipsing binaries. A clear vertical ridge at 1 $d^{-1}$ is likely a light curve systematic. For p-mode pulsators (bottom plot, in frequency), no structures other than the dominant mode can be seen, similar to what was found by \cite{fritzewski2025a}.

\begin{figure}
    \centering
    \includegraphics[width=\linewidth]{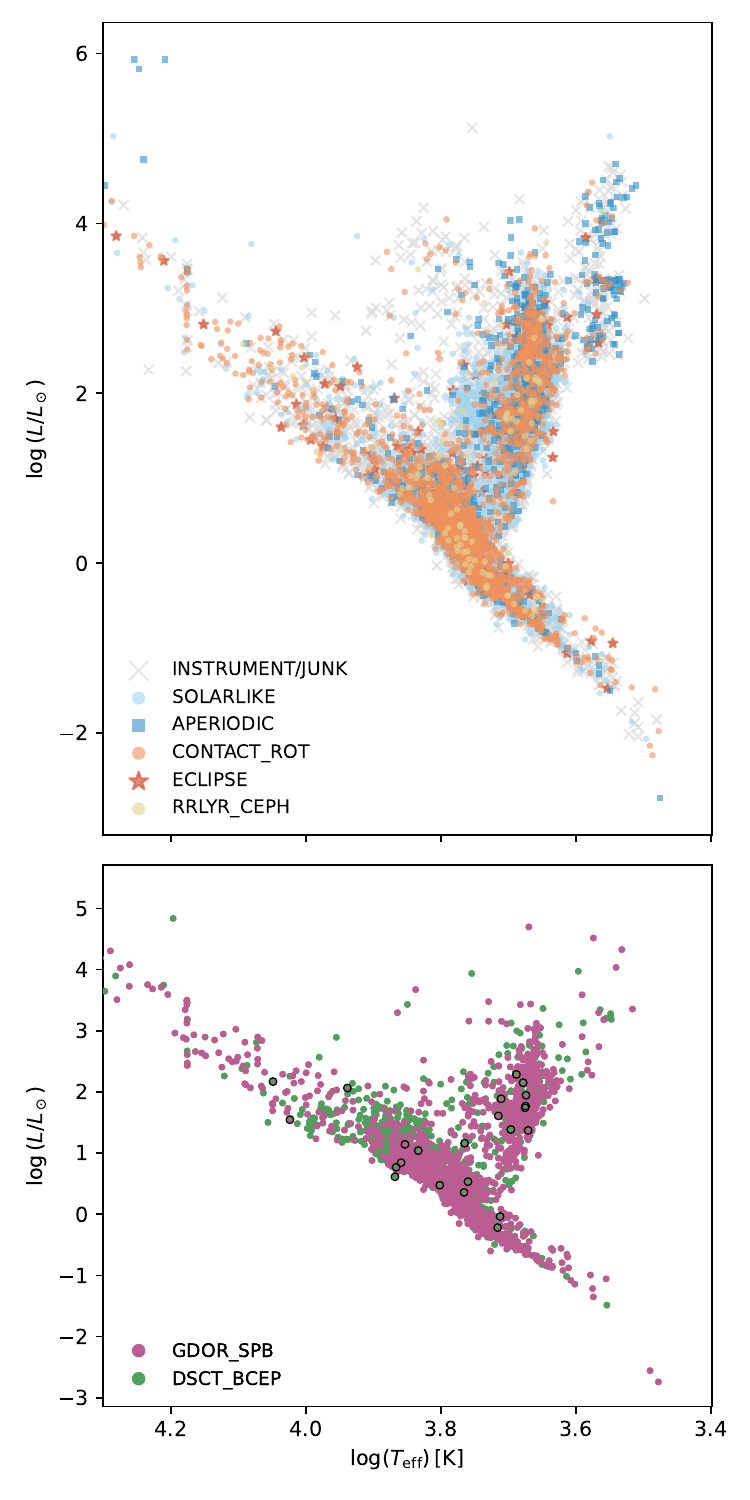}
    \caption{HR Diagram of randomly-sampled high-probability candidate light curves for all classes except DSCT\_BCEP and GDOR\_SPB (top panel) and, separately, only DSCT\_BCEP and GDOR\_SPB with normalized probabilities for the two classes (bottom panel). Black outlines mark stars with the secondary class having a normalized probability higher than 0.2 — potentially hybrid pulsators. A vertical line at 15,000 K is a \textit{Gaia} DR3 grid systematic.}
    \label{fig:hrd}
\end{figure}

Finally, we also investigated the position of candidate pulsators on the Hertzsprung–Russell (HR) diagram. On the top panel of Fig. \ref{fig:hrd}, we show the positions of 5\% randomly-sampled light curves with probabilities higher than 0.5 on the HR diagram (for APERIODIC, ECLIPSE, and INSTRUMENT/JUNK only 1\% is plotted for visibility), which we found to be mostly in line with what expected from the respective types of stars \citep{Aerts2021}. We note that stars classified as SOLARLIKE are found both on the Main Sequence (MS) and in the Giant Branch despite the differences in both excitation mechanisms and typical amplitudes. We manually inspected some light curves for this class in both regions of the HR diagram, which revealed that they share similar light curve and periodogram structures, as expected from stars being put in the same class. We found that some of them lack power excess in frequency ranges expected from either solar-like stars on the MS or solar-like oscillators (pulsating red giants). Particularly, stars labelled SOLARLIKE on the MS where most of the power is concentrated in frequencies below 0.5 $d^{-1}$, are likely misclassifications. This suggests that automatic detection of solar-like oscillators in TESS data is challenging.

The bottom panel shows candidate p- and g-mode pulsators (each point is a normalized probability distribution of a target assigned DSCT\_BCEP and GDOR\_SPB labels), which reveals a number of stars populating space in the gap between $\beta$ Cep / $\delta$ Sct stars and SPB / $\gamma$ Dor stars, similar to \cite{de2023}, \cite{Hey2024}, \cite{mombarg2024}, \cite{aerts2025}, and \cite{Kliapets2025}. Previous studies suggested that these stars could potentially appear cooler due to rotating spots \citep{de2023}. These candidate pulsators are excellent targets for more detailed studies challenging the theoretical bounds of instability strips. We additionally note that a number of stars with high probabilities of being g-mode (and to the lesser extent, p-mode) pulsators, are found in the Red Giant Branch. These are potentially misclassified red giants (solar-like oscillators), which is common for automated pipelines. We tested this hypothesis by inspecting some of these light curves and found that stars labelled GDOR\_SPB fall into one of the two categories: (i) true g-mode pulsators with a wrong $T_{\mathrm{eff}}$; or (ii) predominantly misclassified red giants or, rarer, rotational variables. Stars labelled DSCT\_BCEP are practically entirely true p-mode pulsators with a wrong $T_{\mathrm{eff}}$ and some notable instrumental power excess in the low-frequency regime.

The potential misclassifications revealed by these analyses suggest that using a probabilistic cut-off of 0.5 is too optimistic. Based on visual inspections, we therefore suggest using a threshold per class of 0.75-0.8, depending on accuracy requirements. We do note that even for higher probability bins, we still see some of the discussed confusion happening, with the biggest difference between the confusion matrix in Fig.~\ref{fig:final_conf} and the deployment results happening for the RRLYR\_CEPH class because of the limited training set.

\section{Discussion and conclusions}

In this work, we introduced the ASTRAFier stellar variability classification model. The architecture combines BiLSTM, Attention, and CNN components, which each play an important and complementary role in processing the light curves. The model works directly on the time series, eliminating the need for feature extraction. We have demonstrated the effectiveness of the model in classifying variability, achieving $94.26\%$ classification accuracy on \textit{Kepler} and $88.22\%$ on TESS data. The classification performance is in line with \citet{Audenaert2021} but comes with a much lower computational complexity and model complexity. The rapid classification inference time allows us to more easily classify millions of TESS light curves, while the simpler model architecture allows for better software maintenance. Our deep learning architecture also offers more flexibility for detecting variability classes that are currently not included in our classification scheme, because it does not rely on specific features but can just be retrained to look for new types of patterns.

We found that the performance of our model clearly scales with the size of the training set. Given that Transformers inherently operate within a large hypothesis space, they are known to be particularly data-hungry when trained from scratch, meaning they require large amounts of training data. Although we attempt to mitigate this challenge through the inclusion of LSTM and CNN layers, as well as various regularization techniques, the model remains susceptible to overfitting due to the relatively small size of our labeled training set.

In particular, we can increase the size of our training set by including the data from the TESS extended missions, as we currently only included primary mission data. However, the shorter cadence of the extended mission light curves leads to longer sequences with more time-steps. While this could lead to a more precise light curve with more distinct variability, especially for p-mode pulsators, it is possible that the longer sequences make it more difficult for the model to learn long-range dependencies and increase computational costs. This could potentially be addressed by downsampling the data. For example, \cite{Kliapets2025} found that the recovery of dominant and secondary variability from Kepler in TESS is better in downsampled extended mission data than for the nominal mission data with the same cadence.

We demonstrated the current computational scalability of our approach by classifying $\sim 2.8$ million light curves from TESS sectors 14, 15, and 26, constructing a comprehensive catalog of candidate variable stars in these sectors. The code and trained model are publicly available\footnote{\url{https://github.com/jeraud/TESS-Transformer}}. We are now working on extending our methodology to run on the TESS-Gaia light curves \citep[TGLC,][]{Han2023}, of which the aperture light curve methodology has been incorporated in the QLP pipeline since sector 94\footnote{\url{https://tess.mit.edu/qlp/}} \citep{QLPPetitpas2026}. In particular, we are classifying all TGLC light curves in the PLATO Field-of-View in order to construct a variability catalog for the PLATO Complementary Science Program (Kliapets et al, in prep.).

Lastly, the scaling of data size and performance cannot only be addressed by increasing the size of the labeled training set, but can also be tackled by moving to a self-supervised learning scheme that can take advantage of unlabeled data \citep[see e.g.,][for an explanation]{Parker2024,Audenaert2025}. ASTRAFier is being used to create a foundation model \citep[see e.g.,][for an explanation]{bommasani2021} for TESS \citep{Audenaert2025b} that can be used for a much wider variety of downstream tasks (clustering, anomaly detection, parameter estimation,...), where we are additionally incorporating the ability to remove instrumental and systematic effects \citep{Audenaert2025b,Mercader2026}.


\begin{acknowledgments}
Funding for the TESS, Kepler and K2 mission is provided by NASA's Science Mission Directorate. 
The research leading to these results has received funding from MIT’s Undergraduate Research Opportunities Program (UROP), the BELgian federal Science Policy Office (BELSPO) through the PRODEX grant for PLATO. MK acknowledges The Kavli Foundation for their financial support in the framework of the Kavli Scholarship given to MK from 25/9/2023-24/9/2025, including facilitation of MK's research visit to the MIT Kavli Institute for Astrophysics and Space Research in the fall of 2025 (hosts: JA and GRR). The authors acknowledge the MIT Office of Research Computing and Data (ORCD) for providing high performance computing resources. The authors would like to acknowledge the valuable contributions and feedback provided by members of the TESS Asteroseismic Science Consortium.

\end{acknowledgments}

%

\vspace{5mm}


 \software{astropy \citep{2013A&A...558A..33A, 2018AJ....156..123A}, Lightkurve \citep{lightkurve}, Matplotlib \citep{Hunter:2007}, NumPy \citep{harris2020array}, pandas \citep{mckinney-proc-scipy-2010}, PyTorch \citep{NEURIPS2019_9015}, PyTorch Lightning \citep{Falcon_PyTorch_Lightning_2019}, scikit-learn \citep{scikit-learn}, SciPy \citep{2020SciPy-NMeth}, UMAP \citep{umap}
          }




\bibliography{sample631}{}

@article{aerts2025,
  title={Evolution of the near-core rotation frequency of 2497 intermediate-mass stars from their dominant gravito-inertial mode},
  author={Aerts, Conny and Van Reeth, Timothy and Mombarg, Joey SG and Hey, Daniel},
  journal={Astronomy \& Astrophysics},
  volume={695},
  pages={A214},
  year={2025},
  publisher={EDP Sciences}
}

@article{colman2024,
  title={Methods for the detection of stellar rotation periods in individual TESS sectors and results from the Prime mission},
  author={Colman, Isabel L and Angus, Ruth and David, Trevor and Curtis, Jason and Hattori, Soichiro and Lu, Yuxi Lucy},
  journal={The Astronomical Journal},
  volume={167},
  number={5},
  pages={189},
  year={2024},
  publisher={IOP Publishing}
}

@article{li2020,
  title={Gravity-mode period spacings and near-core rotation rates of 611 $\gamma$ Doradus stars with Kepler},
  author={Li, Gang and Van Reeth, Timothy and Bedding, Timothy R and Murphy, Simon J and Antoci, Victoria and Ouazzani, Rhita-Maria and Barbara, Nicholas H},
  journal={Monthly Notices of the Royal Astronomical Society},
  volume={491},
  number={3},
  pages={3586--3605},
  year={2020},
  publisher={Oxford University Press}
}

@article{fritzewski2025a,
  title={Mode identification and ensemble asteroseismology of 119 $\beta$ Cep stars detected by Gaia light curves and monitored by TESS},
  author={Fritzewski, DJ and Vanrespaille, M and Aerts, C and Guo, Z and Hey, D and De Ridder, J},
  journal={Astronomy \& Astrophysics},
  volume={698},
  pages={A253},
  year={2025},
  publisher={EDP Sciences}
}

@article{fritzewski2025b,
  title={Probing stellar rotation in the Pleiades with gravity-mode pulsators},
  author={Fritzewski, DJ and Kemp, A and Li, G and Aerts, C},
  journal={arXiv preprint arXiv:2512.09395},
  year={2025}
}

@article{de2023,
  title={Gaia Data Release 3-Pulsations in main sequence OBAF-type stars},
  author={De Ridder, Joris and Ripepi, Vincenzo and Aerts, Conny and Palaversa, Lovro and Eyer, Laurent and Holl, Berry and Audard, M and Rimoldini, L and Brown, Anthony GA and Vallenari, Antonella and others},
  journal={Astronomy \& Astrophysics},
  volume={674},
  pages={A36},
  year={2023},
  publisher={EDP sciences}
}

@article{mombarg2024,
  title={Estimates of (convective core) masses, radii, and relative ages for~ 14 000 Gaia-discovered gravity-mode pulsators monitored by TESS},
  author={Mombarg, Joey SG and Aerts, Conny and Van Reeth, Timothy and Hey, Daniel},
  journal={Astronomy \& Astrophysics},
  volume={691},
  pages={A131},
  year={2024},
  publisher={EDP Sciences}
}

@article{aerts2024,
  title={Asteroseismic modelling of fast rotators and its opportunities for astrophysics},
  author={Aerts, Conny and Tkachenko, Andrew},
  journal={Astronomy \& Astrophysics},
  volume={692},
  pages={R1},
  year={2024},
  publisher={EDP Sciences}
}

@ARTICLE{Cui2024,
       author = {{Cui}, Kaiming and {Armstrong}, D.~J. and {Feng}, Fabo},
        title = "{Identifying Light-curve Signals with a Deep-learning-based Object Detection Algorithm. II. A General Light-curve Classification Framework}",
      journal = {\apjs},
         year = 2024,
        month = oct,
       volume = {274},
       number = {2},
          eid = {29},
        pages = {29},
          doi = {10.3847/1538-4365/ad62fd},
archivePrefix = {arXiv},
       eprint = {2311.08080},
 primaryClass = {astro-ph.IM},
       adsurl = {https://ui.adsabs.harvard.edu/abs/2024ApJS..274...29C}
}

@ARTICLE{Audenaert2021,
       author = {{Audenaert}, J. and {Kuszlewicz}, J.~S. and {Handberg}, R. and {Tkachenko}, A. and {Armstrong}, D.~J. and {Hon}, M. and {Kgoadi}, R. and {Lund}, M.~N. and {Bell}, K.~J. and {Bugnet}, L. and {Bowman}, D.~M. and {Johnston}, C. and {Garc{\'\i}a}, R.~A. and {Stello}, D. and {Moln{\'a}r}, L. and {Plachy}, E. and {Buzasi}, D. and {Aerts}, C. and {T'DA collaboration}},
        title = "{TESS Data for Asteroseismology (T'DA) Stellar Variability Classification Pipeline: Setup and Application to the Kepler Q9 Data}",
      journal = {\aj},
         year = 2021,
        month = nov,
       volume = {162},
       number = {5},
          eid = {209},
        pages = {209},
          doi = {10.3847/1538-3881/ac166a},
archivePrefix = {arXiv},
       eprint = {2107.06301},
 primaryClass = {astro-ph.SR},
       adsurl = {https://ui.adsabs.harvard.edu/abs/2021AJ....162..209A}
}

@ARTICLE{Audenaert2022,
       author = {{Audenaert}, J. and {Tkachenko}, A.},
        title = "{Multiscale entropy analysis of astronomical time series. Discovering subclusters of hybrid pulsators}",
      journal = {\aap},
         year = 2022,
        month = oct,
       volume = {666},
          eid = {A76},
        pages = {A76},
          doi = {10.1051/0004-6361/202243469},
archivePrefix = {arXiv},
       eprint = {2206.13529},
 primaryClass = {astro-ph.SR},
       adsurl = {https://ui.adsabs.harvard.edu/abs/2022A&A...666A..76A}
}

@ARTICLE{Barbara2022,
       author = {{Barbara}, Nicholas H. and {Bedding}, Timothy R. and {Fulcher}, Ben D. and {Murphy}, Simon J. and {Van Reeth}, Timothy},
        title = "{Classifying Kepler light curves for 12 000 A and F stars using supervised feature-based machine learning}",
      journal = {\mnras},
         year = 2022,
        month = aug,
       volume = {514},
       number = {2},
        pages = {2793-2804},
          doi = {10.1093/mnras/stac1515},
archivePrefix = {arXiv},
       eprint = {2205.03020},
 primaryClass = {astro-ph.SR},
       adsurl = {https://ui.adsabs.harvard.edu/abs/2022MNRAS.514.2793B}
}

@ARTICLE{Fetherolf2023,
       author = {{Fetherolf}, Tara and {Pepper}, Joshua and {Simpson}, Emilie and {Kane}, Stephen R. and {Mo{\v{c}}nik}, Teo and {English}, John Edward and {Antoci}, Victoria and {Huber}, Daniel and {Jenkins}, Jon M. and {Stassun}, Keivan and {Twicken}, Joseph D. and {Vanderspek}, Roland and {Winn}, Joshua N.},
        title = "{Variability Catalog of Stars Observed during the TESS Prime Mission}",
      journal = {\apjs},
         year = 2023,
        month = sep,
       volume = {268},
       number = {1},
          eid = {4},
        pages = {4},
          doi = {10.3847/1538-4365/acdee5},
archivePrefix = {arXiv},
       eprint = {2208.11721},
 primaryClass = {astro-ph.SR},
       adsurl = {https://ui.adsabs.harvard.edu/abs/2023ApJS..268....4F}
}

@ARTICLE{Hatt2023,
       author = {{Hatt}, Emily and {Nielsen}, Martin B. and {Chaplin}, William J. and {Ball}, Warrick H. and {Davies}, Guy R. and {Bedding}, Timothy R. and {Buzasi}, Derek L. and {Chontos}, Ashley and {Huber}, Daniel and {Kayhan}, Cenk and {Li}, Yaguang and {White}, Timothy R. and {Cheng}, Chen and {Metcalfe}, Travis S. and {Stello}, Dennis},
        title = "{Catalogue of solar-like oscillators observed by TESS in 120-s and 20-s cadence}",
      journal = {\aap},
         year = 2023,
        month = jan,
       volume = {669},
          eid = {A67},
        pages = {A67},
          doi = {10.1051/0004-6361/202244579},
archivePrefix = {arXiv},
       eprint = {2210.09109},
 primaryClass = {astro-ph.SR},
       adsurl = {https://ui.adsabs.harvard.edu/abs/2023A&A...669A..67H}
}

@ARTICLE{Nielsen2022,
       author = {{Nielsen}, M.~B. and {Hatt}, E. and {Chaplin}, W.~J. and {Ball}, W.~H. and {Davies}, G.~R.},
        title = "{A probabilistic method for detecting solar-like oscillations using meaningful prior information. Application to TESS 2-minute photometry}",
      journal = {\aap},
         year = 2022,
        month = jul,
       volume = {663},
          eid = {A51},
        pages = {A51},
          doi = {10.1051/0004-6361/202243064},
archivePrefix = {arXiv},
       eprint = {2203.09404},
 primaryClass = {astro-ph.SR},
       adsurl = {https://ui.adsabs.harvard.edu/abs/2022A&A...663A..51N}
}

@ARTICLE{IJspeert2024b,
       author = {{IJspeert}, L.~W. and {Tkachenko}, A. and {Johnston}, C. and {Aerts}, C.},
        title = "{Statistical view of orbital circularisation with 14 000 characterised TESS eclipsing binaries}",
      journal = {arXiv e-prints},
         year = 2024,
        month = sep,
          eid = {arXiv:2409.20540},
        pages = {arXiv:2409.20540},
          doi = {10.48550/arXiv.2409.20540},
archivePrefix = {arXiv},
       eprint = {2409.20540},
 primaryClass = {astro-ph.SR},
       adsurl = {https://ui.adsabs.harvard.edu/abs/2024arXiv240920540I}
}

@ARTICLE{IJspeert2024a,
       author = {{IJspeert}, L.~W. and {Tkachenko}, A. and {Johnston}, C. and {Pr{\v{s}}a}, A. and {Wells}, M.~A. and {Aerts}, C.},
        title = "{Automated eccentricity measurement from raw eclipsing binary light curves with intrinsic variability}",
      journal = {\aap},
         year = 2024,
        month = may,
       volume = {685},
          eid = {A62},
        pages = {A62},
          doi = {10.1051/0004-6361/202349079},
archivePrefix = {arXiv},
       eprint = {2402.06084},
 primaryClass = {astro-ph.IM},
       adsurl = {https://ui.adsabs.harvard.edu/abs/2024A&A...685A..62I}
}

@ARTICLE{IJspeert2021,
       author = {{IJspeert}, L.~W. and {Tkachenko}, A. and {Johnston}, C. and {Garcia}, S. and {De Ridder}, J. and {Van Reeth}, T. and {Aerts}, C.},
        title = "{An all-sky sample of intermediate- to high-mass OBA-type eclipsing binaries observed by TESS}",
      journal = {\aap},
         year = 2021,
        month = aug,
       volume = {652},
          eid = {A120},
        pages = {A120},
          doi = {10.1051/0004-6361/202141489},
archivePrefix = {arXiv},
       eprint = {2107.10005},
 primaryClass = {astro-ph.SR},
       adsurl = {https://ui.adsabs.harvard.edu/abs/2021A&A...652A.120I}
}

@ARTICLE{Skarka2024,
       author = {{Skarka}, M. and {Henzl}, Z.},
        title = "{Periodic variable A-F spectral type stars in the southern TESS continuous viewing zone. I. Identification and classification}",
      journal = {\aap},
         year = 2024,
        month = aug,
       volume = {688},
          eid = {A25},
        pages = {A25},
          doi = {10.1051/0004-6361/202450711},
archivePrefix = {arXiv},
       eprint = {2406.12578},
 primaryClass = {astro-ph.SR},
       adsurl = {https://ui.adsabs.harvard.edu/abs/2024A&A...688A..25S}
}

@ARTICLE{Skarka2022,
       author = {{Skarka}, M. and {{\v{Z}}{\'a}k}, J. and {Fedurco}, M. and {Paunzen}, E. and {Henzl}, Z. and {Ma{\v{s}}ek}, M. and {Karjalainen}, R. and {Sanchez Arias}, J.~P. and {S{\'o}dor}, {\'A}. and {Auer}, R.~F. and {Kab{\'a}th}, P. and {Karjalainen}, M. and {Li{\v{s}}ka}, J. and {{\v{S}}tegner}, D.},
        title = "{Periodic variable A-F spectral type stars in the northern TESS continuous viewing zone. I. Identification and classification}",
      journal = {\aap},
         year = 2022,
        month = oct,
       volume = {666},
          eid = {A142},
        pages = {A142},
          doi = {10.1051/0004-6361/202244037},
archivePrefix = {arXiv},
       eprint = {2207.12922},
 primaryClass = {astro-ph.SR},
       adsurl = {https://ui.adsabs.harvard.edu/abs/2022A&A...666A.142S}
}

@ARTICLE{Ricker2015,
       author = {{Ricker}, George R. and {Winn}, Joshua N. and {Vanderspek}, Roland and {Latham}, David W. and {Bakos}, G{\'a}sp{\'a}r {\'A}. and {Bean}, Jacob L. and {Berta-Thompson}, Zachory K. and {Brown}, Timothy M. and {Buchhave}, Lars and {Butler}, Nathaniel R. and {Butler}, R. Paul and {Chaplin}, William J. and {Charbonneau}, David and {Christensen-Dalsgaard}, J{\o}rgen and {Clampin}, Mark and {Deming}, Drake and {Doty}, John and {De Lee}, Nathan and {Dressing}, Courtney and {Dunham}, Edward W. and {Endl}, Michael and {Fressin}, Francois and {Ge}, Jian and {Henning}, Thomas and {Holman}, Matthew J. and {Howard}, Andrew W. and {Ida}, Shigeru and {Jenkins}, Jon M. and {Jernigan}, Garrett and {Johnson}, John Asher and {Kaltenegger}, Lisa and {Kawai}, Nobuyuki and {Kjeldsen}, Hans and {Laughlin}, Gregory and {Levine}, Alan M. and {Lin}, Douglas and {Lissauer}, Jack J. and {MacQueen}, Phillip and {Marcy}, Geoffrey and {McCullough}, Peter R. and {Morton}, Timothy D. and {Narita}, Norio and {Paegert}, Martin and {Palle}, Enric and {Pepe}, Francesco and {Pepper}, Joshua and {Quirrenbach}, Andreas and {Rinehart}, Stephen A. and {Sasselov}, Dimitar and {Sato}, Bun'ei and {Seager}, Sara and {Sozzetti}, Alessandro and {Stassun}, Keivan G. and {Sullivan}, Peter and {Szentgyorgyi}, Andrew and {Torres}, Guillermo and {Udry}, Stephane and {Villasenor}, Joel},
        title = "{Transiting Exoplanet Survey Satellite (TESS)}",
      journal = {Journal of Astronomical Telescopes, Instruments, and Systems},
         year = 2015,
        month = jan,
       volume = {1},
          eid = {014003},
        pages = {014003},
          doi = {10.1117/1.JATIS.1.1.014003},
       adsurl = {https://ui.adsabs.harvard.edu/abs/2015JATIS...1a4003R}
}

@ARTICLE{Borucki2010,
       author = {{Borucki}, William J. and {Koch}, David and {Basri}, Gibor and {Batalha}, Natalie and {Brown}, Timothy and {Caldwell}, Douglas and {Caldwell}, John and {Christensen-Dalsgaard}, J{\o}rgen and {Cochran}, William D. and {DeVore}, Edna and {Dunham}, Edward W. and {Dupree}, Andrea K. and {Gautier}, Thomas N. and {Geary}, John C. and {Gilliland}, Ronald and {Gould}, Alan and {Howell}, Steve B. and {Jenkins}, Jon M. and {Kondo}, Yoji and {Latham}, David W. and {Marcy}, Geoffrey W. and {Meibom}, S{\o}ren and {Kjeldsen}, Hans and {Lissauer}, Jack J. and {Monet}, David G. and {Morrison}, David and {Sasselov}, Dimitar and {Tarter}, Jill and {Boss}, Alan and {Brownlee}, Don and {Owen}, Toby and {Buzasi}, Derek and {Charbonneau}, David and {Doyle}, Laurance and {Fortney}, Jonathan and {Ford}, Eric B. and {Holman}, Matthew J. and {Seager}, Sara and {Steffen}, Jason H. and {Welsh}, William F. and {Rowe}, Jason and {Anderson}, Howard and {Buchhave}, Lars and {Ciardi}, David and {Walkowicz}, Lucianne and {Sherry}, William and {Horch}, Elliott and {Isaacson}, Howard and {Everett}, Mark E. and {Fischer}, Debra and {Torres}, Guillermo and {Johnson}, John Asher and {Endl}, Michael and {MacQueen}, Phillip and {Bryson}, Stephen T. and {Dotson}, Jessie and {Haas}, Michael and {Kolodziejczak}, Jeffrey and {Van Cleve}, Jeffrey and {Chandrasekaran}, Hema and {Twicken}, Joseph D. and {Quintana}, Elisa V. and {Clarke}, Bruce D. and {Allen}, Christopher and {Li}, Jie and {Wu}, Haley and {Tenenbaum}, Peter and {Verner}, Ekaterina and {Bruhweiler}, Frederick and {Barnes}, Jason and {Prsa}, Andrej},
        title = "{Kepler Planet-Detection Mission: Introduction and First Results}",
      journal = {Science},
         year = 2010,
        month = feb,
       volume = {327},
       number = {5968},
        pages = {977},
          doi = {10.1126/science.1185402},
       adsurl = {https://ui.adsabs.harvard.edu/abs/2010Sci...327..977B}
}

@ARTICLE{Koch2010,
       author = {{Koch}, David G. and {Borucki}, William J. and {Basri}, Gibor and {Batalha}, Natalie M. and {Brown}, Timothy M. and {Caldwell}, Douglas and {Christensen-Dalsgaard}, J{\o}rgen and {Cochran}, William D. and {DeVore}, Edna and {Dunham}, Edward W. and {Gautier}, Thomas N., III and {Geary}, John C. and {Gilliland}, Ronald L. and {Gould}, Alan and {Jenkins}, Jon and {Kondo}, Yoji and {Latham}, David W. and {Lissauer}, Jack J. and {Marcy}, Geoffrey and {Monet}, David and {Sasselov}, Dimitar and {Boss}, Alan and {Brownlee}, Donald and {Caldwell}, John and {Dupree}, Andrea K. and {Howell}, Steve B. and {Kjeldsen}, Hans and {Meibom}, S{\o}ren and {Morrison}, David and {Owen}, Tobias and {Reitsema}, Harold and {Tarter}, Jill and {Bryson}, Stephen T. and {Dotson}, Jessie L. and {Gazis}, Paul and {Haas}, Michael R. and {Kolodziejczak}, Jeffrey and {Rowe}, Jason F. and {Van Cleve}, Jeffrey E. and {Allen}, Christopher and {Chandrasekaran}, Hema and {Clarke}, Bruce D. and {Li}, Jie and {Quintana}, Elisa V. and {Tenenbaum}, Peter and {Twicken}, Joseph D. and {Wu}, Hayley},
        title = "{Kepler Mission Design, Realized Photometric Performance, and Early Science}",
      journal = {\apjl},
         year = 2010,
        month = apr,
       volume = {713},
       number = {2},
        pages = {L79-L86},
          doi = {10.1088/2041-8205/713/2/L79},
archivePrefix = {arXiv},
       eprint = {1001.0268},
 primaryClass = {astro-ph.EP},
       adsurl = {https://ui.adsabs.harvard.edu/abs/2010ApJ...713L..79K}
}

@ARTICLE{Howell2014,
       author = {{Howell}, Steve B. and {Sobeck}, Charlie and {Haas}, Michael and {Still}, Martin and {Barclay}, Thomas and {Mullally}, Fergal and {Troeltzsch}, John and {Aigrain}, Suzanne and {Bryson}, Stephen T. and {Caldwell}, Doug and {Chaplin}, William J. and {Cochran}, William D. and {Huber}, Daniel and {Marcy}, Geoffrey W. and {Miglio}, Andrea and {Najita}, Joan R. and {Smith}, Marcie and {Twicken}, J.~D. and {Fortney}, Jonathan J.},
        title = "{The K2 Mission: Characterization and Early Results}",
      journal = {\pasp},
         year = 2014,
        month = apr,
       volume = {126},
       number = {938},
        pages = {398},
          doi = {10.1086/676406},
archivePrefix = {arXiv},
       eprint = {1402.5163},
 primaryClass = {astro-ph.IM},
       adsurl = {https://ui.adsabs.harvard.edu/abs/2014PASP..126..398H}
}

@ARTICLE{Hon2018a,
       author = {{Hon}, Marc and {Stello}, Dennis and {Zinn}, Joel C.},
        title = "{Detecting Solar-like Oscillations in Red Giants with Deep Learning}",
      journal = {\apj},
         year = 2018,
        month = may,
       volume = {859},
       number = {1},
          eid = {64},
        pages = {64},
          doi = {10.3847/1538-4357/aabfdb},
archivePrefix = {arXiv},
       eprint = {1804.07495},
 primaryClass = {astro-ph.SR},
       adsurl = {https://ui.adsabs.harvard.edu/abs/2018ApJ...859...64H}
}

@ARTICLE{Hon2018b,
       author = {{Hon}, Marc and {Stello}, Dennis and {Yu}, Jie},
        title = "{Deep learning classification in asteroseismology using an improved neural network: results on 15 000 Kepler red giants and applications to K2 and TESS data}",
      journal = {\mnras},
         year = 2018,
        month = may,
       volume = {476},
       number = {3},
        pages = {3233-3244},
          doi = {10.1093/mnras/sty483},
archivePrefix = {arXiv},
       eprint = {1802.07260},
 primaryClass = {astro-ph.IM},
       adsurl = {https://ui.adsabs.harvard.edu/abs/2018MNRAS.476.3233H}
}

@ARTICLE{Hon2019,
       author = {{Hon}, Marc and {Stello}, Dennis and {Garc{\'\i}a}, Rafael A. and {Mathur}, Savita and {Sharma}, Sanjib and {Colman}, Isabel L. and {Bugnet}, Lisa},
        title = "{A search for red giant solar-like oscillations in all Kepler data}",
      journal = {\mnras},
         year = 2019,
        month = jun,
       volume = {485},
       number = {4},
        pages = {5616-5630},
          doi = {10.1093/mnras/stz622},
archivePrefix = {arXiv},
       eprint = {1903.00115},
 primaryClass = {astro-ph.SR},
       adsurl = {https://ui.adsabs.harvard.edu/abs/2019MNRAS.485.5616H}
}

@ARTICLE{Hey2024,
       author = {{Hey}, Daniel and {Aerts}, Conny},
        title = "{Confronting sparse Gaia DR3 photometry with TESS for a sample of around 60 000 OBAF-type pulsators}",
      journal = {\aap},
         year = 2024,
        month = aug,
       volume = {688},
          eid = {A93},
        pages = {A93},
          doi = {10.1051/0004-6361/202450489},
archivePrefix = {arXiv},
       eprint = {2405.01539},
 primaryClass = {astro-ph.SR},
       adsurl = {https://ui.adsabs.harvard.edu/abs/2024A&A...688A..93H}
}

@ARTICLE{Eschen2024,
       author = {{Eschen}, Yoshi Nike Emilia and {Bayliss}, Daniel and {Wilson}, Thomas G. and {Kunimoto}, Michelle and {Pelisoli}, Ingrid and {Rodel}, Toby},
        title = "{Viewing the PLATO LOPS2 Field Through the Lenses of TESS}",
      journal = {arXiv e-prints},
         year = 2024,
        month = sep,
          eid = {arXiv:2409.13039},
        pages = {arXiv:2409.13039},
          doi = {10.48550/arXiv.2409.13039},
archivePrefix = {arXiv},
       eprint = {2409.13039},
 primaryClass = {astro-ph.EP},
       adsurl = {https://ui.adsabs.harvard.edu/abs/2024arXiv240913039E}
}

@ARTICLE{Rauer2024,
       author = {{Rauer}, Heike and {Aerts}, Conny and {Cabrera}, Juan and {Deleuil}, Magali and {Erikson}, Anders and {Gizon}, Laurent and {Goupil}, Mariejo and {Heras}, Ana and {Lorenzo-Alvarez}, Jose and {Marliani}, Filippo and {Martin-Garcia}, Cesar and {Mas-Hesse}, J. Miguel and {O'Rourke}, Laurence and {Osborn}, Hugh and {Pagano}, Isabella and {Piotto}, Giampaolo and {Pollacco}, Don and {Ragazzoni}, Roberto and {Ramsay}, Gavin and {Udry}, St{\'e}phane and {Appourchaux}, Thierry and {Benz}, Willy and {Brandeker}, Alexis and {G{\"u}del}, Manuel and {Janot-Pacheco}, Eduardo and {Kabath}, Petr and {Kjeldsen}, Hans and {Min}, Michiel and {Santos}, Nuno and {Smith}, Alan and {Suarez}, Juan-Carlos and {Werner}, Stephanie C. and {Aboudan}, Alessio and {Abreu}, Manuel and {Acu{\~n}a}, Lorena and {Adams}, Moritz and {Adibekyan}, Vardan and {Affer}, Laura and {Agneray}, Fran{\c{c}}ois and {Agnor}, Craig and {Aguirre B{\o}rsen-Koch}, Victor and {Ahmed}, Saad and {Aigrain}, Suzanne and {Al-Bahlawan}, Ashraf and {Alcacera Gil}, M de los Angeles and {Alei}, Eleonora and {Alencar}, Silvia and {Alexander}, Richard and {Alfonso-Garz{\'o}n}, Julia and {Alibert}, Yann and {Allende Prieto}, Carlos and {Almeida}, Leonardo and {Alonso Sobrino}, Roi and {Altavilla}, Giuseppe and {Althaus}, Christian and {Alonso Alvarez Trujillo}, Luis and {Amarsi}, Anish and {Ammler-von Eiff}, Matthias and {Am{\^o}res}, Eduardo and {Andrade}, Laerte and {Antoniadis-Karnavas}, Alexandros and {Ant{\'o}nio}, Carlos and {Aparicio del Moral}, Beatriz and {Appolloni}, Matteo and {Arena}, Claudio and {Armstrong}, David and {Aroca Aliaga}, Jose and {Asplund}, Martin and {Audenaert}, Jeroen and {Auricchio}, Natalia and {Avelino}, Pedro and {Baeke}, Ann and {Bailli{\'e}}, Kevin and {Balado}, Ana and {Balestra}, Andrea and {Ball}, Warrick and {Ballans}, Herve and {Ballot}, Jerome and {Barban}, Caroline and {Barbary}, Ga{\"e}le and {Barbieri}, Mauro and {Barcel{\'o} Forteza}, Sebasti{\`a} and {Barker}, Adrian and {Barklem}, Paul and {Barnes}, Sydney and {Barrado Navascues}, David and {Barragan}, Oscar and {Baruteau}, Cl{\'e}ment and {Basu}, Sarbani and {Baudin}, Frederic and {Baumeister}, Philipp and {Bayliss}, Daniel and {Bazot}, Michael and {Beck}, Paul G. and {Bedding}, Tim and {Belkacem}, Kevin and {Bellinger}, Earl and {Benatti}, Serena and {Benomar}, Othman and {B{\'e}rard}, Diane and {Bergemann}, Maria and {Bergomi}, Maria and {Bernardo}, Pierre and {Biazzo}, Katia and {Bignamini}, Andrea and {Bigot}, Lionel and {Billot}, Nicolas and {Binet}, Martin and {Biondi}, David and {Biondi}, Federico and {Birch}, Aaron C. and {Bitsch}, Bertram and {Bluhm Ceballos}, Paz Victoria and {B{\'o}di}, Attila and {Bogn{\'a}r}, Zs{\'o}fia and {Boisse}, Isabelle and {Bolmont}, Emeline and {Bonanno}, Alfio and {Bonavita}, Mariangela and {Bonfanti}, Andrea and {Bonfils}, Xavier and {Bonito}, Rosaria and {Bonomo}, Aldo Stefano and {B{\"o}rner}, Anko and {Boro Saikia}, Sudeshna and {Borreguero Mart{\'\i}n}, Elisa and {Borsa}, Francesco and {Borsato}, Luca and {Bossini}, Diego and {Bouchy}, Francois and {Bou{\'e}}, Gwena{\"e}l and {Boufleur}, Rodrigo and {Boumier}, Patrick and {Bourrier}, Vincent and {Bowman}, Dominic M. and {Bozzo}, Enrico and {Bradley}, Louisa and {Bray}, John and {Bressan}, Alessandro and {Breton}, Sylvain and {Brienza}, Daniele and {Brito}, Ana and {Brogi}, Matteo and {Brown}, Beverly and {Brown}, David and {Brun}, Allan Sacha and {Bruno}, Giovanni and {Bruns}, Michael and {Buchhave}, Lars A. and {Bugnet}, Lisa and {Buldgen}, Ga{\"e}l and {Burgess}, Patrick and {Busatta}, Andrea and {Busso}, Giorgia and {Buzasi}, Derek and {Caballero}, Jos{\'e} A. and {Cabral}, Alexandre and {Calderone}, Flavia and {Cameron}, Robert and {Cameron}, Andrew and {Campante}, Tiago and {Canto Martins}, Bruno Leonardo and {Cara}, Christophe and {Carone}, Ludmila and {Carrasco}, Josep Manel and {Casagrande}, Luca and {Casewell}, Sarah L. and {Cassisi}, Santi and {Castellani}, Marco and {Castro}, Matthieu and {Catala}, Claude and {Catal{\'a}n Fern{\'a}ndez}, Irene and {Catelan}, M{\'a}rcio and {Cegla}, Heather and {Cerruti}, Chiara and {Cessa}, Virginie and {Chadid}, Merieme and {Chaplin}, William and {Charpinet}, Stephane and {Chiappini}, Cristina and {Chiarucci}, Simone and {Chiavassa}, Andrea and {Chinellato}, Simonetta and {Chirulli}, Giovanni and {Christensen-Dalsgaard}, Jorgen and {Church}, Ross and {Claret}, Antonio and {Clarke}, Cathie and {Claudi}, Riccardo and {Clermont}, Lionel and {Coelho}, Hugo and {Coelho}, Joao and {Cogato}, Fabrizio and {Colom{\'e}}, Josep and {Condamin}, Mathieu and {Conseil}, Simon and {Corbard}, Thierry and {Correia}, Alexandre C.~M. and {Corsaro}, Enrico and {Cosentino}, Rosario and {Costes}, Jean and {Cottinelli}, Andrea and {Covone}, Giovanni and {Creevey}, Orlagh L. and {Crida}, Aurelien and {Csizmadia}, Szilard and {Cunha}, Margarida and {Curry}, Patrick and {da Costa}, Jefferson and {da Silva}, Francys and {Dalal}, Shweta and {Damasso}, Mario and {Damiani}, Cilia and {Damiani}, Francesco and {Liduina das Chagas}, Maria and {Davies}, Melvyn and {Davies}, Guy and {Davies}, Ben and {Davison}, Gary and {de Almeida}, Leandro and {de Angeli}, Francesca and {Cabral de Barros}, Susana Cristina and {de Castro Le{\~a}o}, Izan and {Brito de Freitas}, Daniel and {de Freitas}, Marcia Cristina and {De Martino}, Domitilla and {Renan de Medeiros}, Jos{\'e} and {de Paula}, Luiz Alberto and {de Plaa}, Jelle and {De Ridder}, Joris and {Deal}, Morgan and {Decin}, Leen and {Deeg}, Hans and {Degl'Innocenti}, Scilla and {Deheuvels}, Sebastien and {del Burgo}, Carlos and {Del Sordo}, Fabio and {Delgado-Mena}, Elisa and {Demangeon}, Olivier and {Denk}, Tilmann and {Derekas}, Aliz and {Desidera}, Silvano and {Dexet}, Marc and {Di Criscienzo}, Marcella and {Di Giorgio}, Anna Maria and {Di Mauro}, Maria Pia and {Diaz Rial}, Federico Jose and {D{\'\i}az-Garc{\'\i}a}, Jos{\'e}-Javier and {Dima}, Marco and {Dinuzzi}, Giacomo and {Dionatos}, Odysseas and {Distefano}, Elisa and {do Nascimento}, Jose-Dias, Jr. and {Domingo}, Albert and {D'Orazi}, Valentina and {Dorn}, Caroline and {Doyle}, Lauren and {Duarte}, Elena and {Ducellier}, Florent and {Dumaye}, Luc and {Dumusque}, Xavier and {Dupret}, Marc-Antoine and {Eggenberger}, Patrick and {Ehrenreich}, David and {Eigm{\"u}ller}, Philipp and {Eising}, Johannes and {Emilio}, Marcelo and {Eriksson}, Kjell and {Ermocida}, Marco and {Isidoro Escate Giribaldi}, Riano and {Eschen}, Yoshi and {Estrela}, In{\^e}s and {Evans}, Dafydd Wyn and {Fabbian}, Damian and {Fabrizio}, Michele and {Faria}, Jo{\~a}o Pedro and {Farina}, Maria and {Farinato}, Jacopo and {Feliz}, Dax and {Feltzing}, Sofia and {Fenouillet}, Thomas and {Ferrari}, Lorenza and {Ferraz-Mello}, Sylvio and {Fialho}, Fabio and {Fienga}, Agnes and {Figueira}, Pedro and {Fiori}, Laura and {Flaccomio}, Ettore and {Focardi}, Mauro and {Foley}, Steve and {Fontignie}, Jean and {Ford}, Dominic and {Fornazier}, Karin and {Forveille}, Thierry and {Fossati}, Luca and {de Marca Franca}, Rodrigo and {da Silva}, Lucas Franco and {Frasca}, Antonio and {Fridlund}, Malcolm and {Furlan}, Marco and {Gabler}, Sarah-Maria and {Gaido}, Marco and {Gallagher}, Andrew and {Galli}, Emanuele and {Garcia}, Rafael A. and {Garc{\'\i}a Hern{\'a}ndez}, Antonio and {Garcia Munoz}, Antonio and {Garc{\'\i}a-V{\'a}zquez}, Hugo and {Garrido Haba}, Rafael and {Gaulme}, Patrick and {Gauthier}, Nicolas and {Gehan}, Charlotte and {Gent}, Matthew and {Georgieva}, Iskra and {Ghigo}, Mauro and {Giana}, Edoardo and {Gill}, Samuel and {Girardi}, Leo and {Giuliatti Winter}, Silvia and {Giusi}, Giovanni and {Gomes da Silva}, Jo{\~a}o and {G{\'o}mez Zazo}, Luis Jorge and {Gomez-Lopez}, Juan Manuel and {Isai Gonz{\'a}lez Hern{\'a}ndez}, Jonay and {Gonzalez Murillo}, Kevin and {Gorius}, Nicolas and {Gouel}, Pierre-Vincent and {Goulty}, Duncan and {Granata}, Valentina and {Grenfell}, John Lee and {Grie{\ss}bach}, Denis and {Grolleau}, Emmanuel and {Grouffal}, Salom{\'e} and {Grziwa}, Sascha and {Guarcello}, Mario Giuseppe and {Gueguen}, Lo{\"\i}c and {Guenther}, Eike Wolf and {Guilhem}, Terrasa and {Guillerot}, Lucas and {Guiot}, Pierre and {Guterman}, Pascal and {Guti{\'e}rrez}, Antonio and {Guti{\'e}rrez-Canales}, Fernando and {Hagelberg}, Janis and {Haldemann}, Jonas and {Hall}, Cassandra and {Handberg}, Rasmus and {Harrison}, Ian and {Harrison}, Diana L. and {Hasiba}, Johann and {Haswell}, Carole A. and {Hatalova}, Petra and {Hatzes}, Artie and {Haywood}, Raphaelle and {H{\'e}brard}, Guillaume and {Heckes}, Frank and {Heiter}, Ulrike and {Hekker}, Saskia and {Heller}, Ren{\'e} and {Helling}, Christiane and {Helminiak}, Krzysztof and {Hemsley}, Simon and {Heng}, Kevin and {Hermans}, Aline and {Hermes}, JJ and {Hidalgo Torres}, Nadia and {Hinkel}, Natalie and {Hobbs}, David and {Hodgkin}, Simon and {Hofmann}, Karl and {Hojjatpanah}, Saeed and {Houdek}, G{\"u}nter and {Huber}, Daniel and {Huesler}, Joseph and {Hui-Bon-Hoa}, Alain and {Huygen}, Rik and {Huynh}, Duc-Dat and {Iro}, Nicolas and {Irwin}, Jonathan and {Irwin}, Mike and {Izidoro}, Andr{\'e} and {Jacquinod}, Sophie and {Emborg Jannsen}, Nicholas and {Janson}, Markus and {Jeszenszky}, Harald and {Jiang}, Chen and {Jos{\'e} Jimenez Mancebo}, Antonio and {Jofre}, Paula and {Johansen}, Anders and {Johnston}, Cole and {Jones}, Geraint and {Kallinger}, Thomas and {K{\'a}lm{\'a}n}, Szil{\'a}rd and {Kanitz}, Thomas and {Karjalainen}, Marie and {Karjalainen}, Raine and {Karoff}, Christoffer and {Kawaler}, Steven and {Kawata}, Daisuke and {Keereman}, Arnoud and {Keiderling}, David and {Kennedy}, Tom and {Kenworthy}, Matthew and {Kerschbaum}, Franz and {Kidger}, Mark and {Kiefer}, Flavien and {Kintziger}, Christian and {Kislyakova}, Kristina and {Kiss}, L{\'a}szl{\'o} and {Klagyivik}, Peter and {Klahr}, Hubert and {Klevas}, Jonas and {Kochukhov}, Oleg and {K{\"o}hler}, Ulrich and {Kolb}, Ulrich and {Koncz}, Alexander and {Korth}, Judith and {Kostogryz}, Nadiia and {Kov{\'a}cs}, G{\'a}bor and {Kov{\'a}cs}, J{\'o}zsef and {Kozhura}, Oleg and {Krivova}, Natalie and {Ku{\v{c}}inskas}, Arunas and {Kuhlemann}, Ilyas and {Kupka}, Friedrich and {Laauwen}, Wouter and {Labiano}, Alvaro and {Lagarde}, Nadege and {Laget}, Philippe and {Laky}, Gunter and {Lam}, Kristine Wai Fun and {Lambrechts}, Michiel and {Lammer}, Helmut and {Lanza}, Antonino Francesco and {Lanzafame}, Alessandro and {Lares Martiz}, Mariel and {Laskar}, Jacques and {Latter}, Henrik and {Lavanant}, Tony and {Lawrenson}, Alastair and {Lazzoni}, Cecilia and {Lebre}, Agnes and {Lebreton}, Yveline and {Lecavelier des Etangs}, Alain and {Leinhardt}, Zoe and {Leleu}, Adrien and {Lendl}, Monika and {Leto}, Giuseppe and {Levillain}, Yves and {Libert}, Anne-Sophie and {Lichtenberg}, Tim and {Ligi}, Roxanne and {Lignieres}, Francois and {Lillo-Box}, Jorge and {Linsky}, Jeffrey and {Scige Liu}, John and {Loidolt}, Dominik and {Longval}, Yuying and {Lopes}, Il{\'\i}dio and {Lorenzani}, Andrea and {Ludwig}, Hans-Guenter and {Lund}, Mikkel and {Sloth Lundkvist}, Mia and {Luri}, Xavier and {Maceroni}, Carla and {Madden}, Sean and {Madhusudhan}, Nikku and {Maggio}, Antonio and {Magliano}, Christian and {Magrin}, Demetrio and {Mahy}, Laurent and {Maibaum}, Olaf and {Malac-Allain}, LeeRoy and {Malapert}, Jean-Christophe and {Malavolta}, Luca and {Maldonado}, Jesus and {Mamonova}, Elena and {Manchon}, Louis and {Mann}, Andrew and {Mantovan}, Giacomo and {Marafatto}, Luca and {Marconi}, Marcella and {Mardling}, Rosemary and {Marigo}, Paola and {Marinoni}, Silvia and {Marques}, {\'E}rico and {Marques}, Joao Pedro and {Marrese}, Paola Maria and {Marshall}, Douglas and {Mart{\'\i}nez Perales}, Silvia and {Mary}, David and {Marzari}, Francesco and {Masana}, Eduard and {Mascher}, Andrina and {Mathis}, St{\'e}phane and {Mathur}, Savita and {Mattiuci Figueiredo}, Ana Carolina and {Maxted}, Pierre F.~L. and {Mazeh}, Tsevi and {Mazevet}, Stephane and {Mazzei}, Francesco and {McCormac}, James and {McMillan}, Paul and {Menou}, Lucas and {Merle}, Thibault and {Meru}, Farzana and {Mesa}, Dino and {Messina}, Sergio and {M{\'e}sz{\'a}ros}, Szabolcs and {Meunier}, Nad{\'e}ge and {Meunier}, Jean-Charles and {Micela}, Giuseppina and {Michaelis}, Harald and {Michel}, Eric and {Michielsen}, Mathias and {Michtchenko}, Tatiana and {Miglio}, Andrea and {Miguel}, Yamila and {Milligan}, David and {Mirouh}, Giovanni and {Mitchell}, Morgan A. and {Moedas}, Nuno and {Molendini}, Francesca and {Moln{\'a}r}, L{\'a}szl{\'o} and {Mombarg}, Joey and {Montalban}, Josefina and {Montalto}, Marco and {Monteiro}, M{\'a}rio J.~P.~F.~G. and {Morales}, Juan Carlos and {Morales-Calderon}, Maria and {Morbidelli}, Alessandro and {Mordasini}, Christoph and {Moreau}, Chrystel and {Morel}, Thierry and {Morello}, Guiseppe and {Morin}, Julien and {Mortier}, Annelies and {Mosser}, Beno{\^\i}t and {Mourard}, Denis and {Mousis}, Olivier and {Moutou}, Claire and {Mowlavi}, Nami and {Moya}, Andr{\'e}s and {Muehlmann}, Prisca and {Muirhead}, Philip and {Munari}, Matteo and {Musella}, Ilaria and {Mustill}, Alexander James and {Nardetto}, Nicolas and {Nardiello}, Domenico and {Narita}, Norio and {Nascimbeni}, Valerio and {Nash}, Anna and {Neiner}, Coralie and {Nelson}, Richard P. and {Nettelmann}, Nadine and {Nicolini}, Gianalfredo and {Nielsen}, Martin and {Niemi}, Sami-Matias and {Noack}, Lena and {Noels-Grotsch}, Arlette and {Noll}, Anthony and {Norazman}, Azib and {Norton}, Andrew J. and {Nsamba}, Benard and {Ofir}, Aviv and {Ogilvie}, Gordon and {Olander}, Terese and {Olivetto}, Christian and {Olofsson}, G{\"o}ran and {Ong}, Joel and {Ortolani}, Sergio and {Oshagh}, Mahmoudreza and {Ottacher}, Harald and {Ottensamer}, Roland and {Ouazzani}, Rhita-Maria and {Paardekooper}, Sijme-Jan and {Pace}, Emanuele and {Pajas}, Miriam and {Palacios}, Ana and {Palandri}, Gaelle and {Palle}, Enric and {Paproth}, Carsten and {Parro}, Vanderlei and {Parviainen}, Hannu and {Granado}, Javier Pascual and {Passegger}, Vera Maria and {Pastor-Morales}, Carmen and {P{\"a}tzold}, Martin and {Gade Pedersen}, May and {Pena Hidalgo}, David and {Pepe}, Francesco and {Pereira}, Filipe and {Persson}, Carina M. and {Pertenais}, Martin and {Peter}, Gisbert and {Petit}, Antoine C. and {Petit}, Pascal and {Pezzuto}, Stefano and {Pichierri}, Gabriele and {Pietrinferni}, Adriano and {Pinheiro}, Fernando and {Pinsonneault}, Marc and {Plachy}, Emese and {Plasson}, Philippe and {Plez}, Bertrand and {Poppenhaeger}, Katja and {Poretti}, Ennio and {Portaluri}, Elisa and {Portell}, Jordi and {Frederico Porto de Mello}, Gustavo and {Poyatos}, Julien and {Pozuelos}, Francisco J. and {Prada Moroni}, Pier Giorgio and {Pricopi}, Dumitru and {Prisinzano}, Loredana and {Quade}, Matthias and {Quirrenbach160}, ndreas and {Rabanal Reina6}, Julio Arturo and {Rabello Soares}, Maria Cristina and {Raimondo}, Gabriella and {Rainer}, Monica and {Ram{\'o}n Rod{\'o}n}, Jose and {Ram{\'o}n-Ballesta}, Alejandro and {Ramos Zapata}, Gonzalo and {R{\"a}tz}, Stefanie and {Rauterberg}, Christoph and {Redman}, Bob and {Redmer}, Ronald and {Reese}, Daniel and {Regibo}, Sara and {Reiners}, Ansgar and {Reinhold}, Timo and {Renie}, Christian and {Ribas}, Ignasi and {Ribeiro}, Sergio and {Pereira Ricciardi}, Thiago and {Rice}, Ken and {Richard}, Olivier and {Riello}, Marco and {Rieutord}, Michel and {Ripepi}, Vincenzo and {Rixon}, Guy and {Rockstein}, Steve and {Rodr{\'\i}guez}, Mar{\'\i}a Teresa Rodrigo and {Rodr{\'\i}guez D{\'\i}az}, Luisa Fernanda and {Rodriguez Garcia}, Juan Pablo and {Rodriguez-Gomez}, Julio and {Roehlly}, Yannick and {Roig}, Fernando and {Rojas-Ayala}, B{\'a}rbara and {Rolf}, Tobias and {Lysgaard R{\o}rsted}, Jakob and {Rosado}, Hugo and {Rosotti}, Giovanni and {Roth}, Olivier and {Roth}, Markus and {Rousseau}, Alex and {Roxburgh}, Ian and {Roy}, Fabrice and {Royer}, Pierre and {Ruane}, Kirk and {Rufini Mastropasqua}, Sergio and {Ruiz de Galarreta}, Claudia and {Russi}, Andrea and {Saar}, Steven and {Saillenfest}, Melaine and {Salaris}, Maurizio and {Salmon}, Sebastien and {Saltas}, Ippocratis and {Samadi}, R{\'e}za and {Samadi}, Aunia and {Samra}, Dominic and {Sanches da Silva}, Tiago and {Andr{\'e}s S{\'a}nchez Carrasco}, Miguel and {Santerne}, Alexandre and {Santoli}, Francesco and {Santos}, {\^A}ngela R.~G. and {Sanz Mesa}, Rosario and {Sarro}, Luis Manuel and {Scandariato}, Gaetano and {Sch{\"a}fer}, Martin and {Schlafly}, Edward and {Schmider}, Fran{\c{c}}ois-Xavier and {Schneider}, Jean and {Schou}, Jesper and {Schunker}, Hannah and {J{\"o}rg Schwarzkopf}, Gabriel and {Serenelli}, Aldo and {Seynaeve}, Dries and {Shan}, Yutong and {Shapiro}, Alexander and {Shipman}, Russel and {Sicilia}, Daniela and {Sierra Sanmartin}, Maria Angeles and {Sigot}, Axelle and {Silliman}, Kyle and {Silvotti}, Roberto and {Simon}, Attila E. and {Simoyama Napoli}, Ricardo and {Skarka}, Marek and {Smalley}, Barry and {Smiljanic}, Rodolfo and {Smit}, Samuel and {Smith}, Alexis and {Smith}, Leigh and {Snellen}, Ignas and {S{\'o}dor}, {\'A}d{\'a}m and {Sohl}, Frank and {Solanki}, Sami K. and {Sortino}, Francesca and {Sousa}, S{\'e}rgio and {Southworth}, John and {Souto}, Diogo and {Sozzetti}, Alessandro and {Stamatellos}, Dimitris and {Stassun}, Keivan and {Steller}, Manfred and {Stello}, Dennis and {Stelzer}, Beate and {Stiebeler}, Ulrike and {Stokholm}, Amalie and {Storelvmo}, Trude and {Strassmeier}, Klaus and {Str{\o}m}, Paul Anthony and {Strugarek}, Antoine and {Sulis}, Sophia and {{\v{S}}vanda}, Michal and {Szabados}, L{\'a}szl{\'o} and {Szab{\'o}}, R{\'o}bert and {Szab{\'o}}, Gyula M. and {Szuszkiewicz}, Ewa and {Talens}, Geert Jan and {Teti}, Daniele and {Theisen}, Tom and {Th{\'e}venin}, Fr{\'e}d{\'e}ric and {Thoul}, Anne and {Tiphene}, Didier and {Titz-Weider}, Ruth and {Tkachenko}, Andrew and {Tomecki}, Daniel and {Tonfat}, Jorge and {Tosi}, Nicola and {Trampedach}, Regner and {Traven}, Gregor and {Triaud}, Amaury and {Tr{\o}nnes}, Reidar and {Tsantaki}, Maria and {Tschentscher}, Matthias and {Turin}, Arnaud and {Tvaruzka}, Adam and {Ulmer}, Bernd and {Ulmer-Moll}, Sol{\`e}ne and {Ulusoy}, Ceren and {Umbriaco}, Gabriele and {Valencia}, Diana and {Valentini}, Marica and {Valio}, Adriana and {Valverde Guijarro}, {\'A}ngel Luis and {Van Eylen}, Vincent and {Van Grootel}, Valerie and {van Kempen}, Tim A. and {Van Reeth}, Timothy and {Van Zelst}, Iris and {Vandenbussche}, Bart and {Vasiliou}, Konstantinos and {Vasilyev}, Valeriy and {Vaz de Mascarenhas}, David and {Vazan}, Allona and {Vela Nunez}, Marina and {Nunes Velloso}, Eduardo and {Ventura}, Rita and {Ventura}, Paolo and {Venturini}, Julia and {Trallero}, Isabel Vera and {Veras}, Dimitri and {Verdugo}, Eva and {Verma}, Kuldeep and {Vibert}, Didier and {Vicanek Martinez}, Tobias and {Vida}, Kriszti{\'a}n and {Vigan}, Arthur and {Villacorta}, Antonio and {Villaver}, Eva and {Villaverde Aparicio}, Marcos and {Viotto}, Valentina and {Vorobyov}, Eduard and {Vorontsov}, Sergey and {Wagner}, Frank W. and {Walloschek}, Thomas and {Walton}, Nicholas and {Walton}, Dave and {Wang}, Haiyang and {Waters}, Rens and {Watson}, Christopher and {Wedemeyer}, Sven and {Weeks}, Angharad and {Weingril}, J{\"o}rg and {Weiss}, Annita and {Wendler}, Belinda and {West}, Richard and {Westerdorff}, Karsten and {Westphal}, Pierre-Amaury and {Wheatley}, Peter and {White}, Tim and {Whittaker}, Amadou and {Wickhusen}, Kai and {Wilson}, Thomas and {Windsor}, James and {Winter}, Othon and {Lykke Winther}, Mark and {Winton}, Alistair and {Witteck}, Ulrike and {Witzke}, Veronika and {Woitke}, Peter and {Wolter}, David and {Wuchterl}, G{\"u}nther and {Wyatt}, Mark and {Yang}, Dan and {Yu}, Jie and {Zanmar Sanchez}, Ricardo and {Rosa Zapatero Osorio}, Mar{\'\i}a and {Zechmeister}, Mathias and {Zhou}, Yixiao and {Ziemke}, Claas and {Zwintz}, Konstanze},
        title = "{The PLATO Mission}",
      journal = {arXiv e-prints},
         year = 2024,
        month = jun,
          eid = {arXiv:2406.05447},
        pages = {arXiv:2406.05447},
          doi = {10.48550/arXiv.2406.05447},
archivePrefix = {arXiv},
       eprint = {2406.05447},
 primaryClass = {astro-ph.IM},
       adsurl = {https://ui.adsabs.harvard.edu/abs/2024arXiv240605447R}
}

@ARTICLE{Armstrong2016,
       author = {{Armstrong}, D.~J. and {Kirk}, J. and {Lam}, K.~W.~F. and {McCormac}, J. and {Osborn}, H.~P. and {Spake}, J. and {Walker}, S. and {Brown}, D.~J.~A. and {Kristiansen}, M.~H. and {Pollacco}, D. and {West}, R. and {Wheatley}, P.~J.},
        title = "{K2 variable catalogue - II. Machine learning classification of variable stars and eclipsing binaries in K2 fields 0-4}",
      journal = {\mnras},
         year = 2016,
        month = feb,
       volume = {456},
       number = {2},
        pages = {2260-2272},
          doi = {10.1093/mnras/stv2836},
archivePrefix = {arXiv},
       eprint = {1512.01246},
 primaryClass = {astro-ph.SR},
       adsurl = {https://ui.adsabs.harvard.edu/abs/2016MNRAS.456.2260A}
}

@ARTICLE{Naul2018,
       author = {{Naul}, Brett and {Bloom}, Joshua S. and {P{\'e}rez}, Fernando and {van der Walt}, St{\'e}fan},
        title = "{A recurrent neural network for classification of unevenly sampled variable stars}",
      journal = {Nature Astronomy},
         year = 2018,
        month = nov,
       volume = {2},
        pages = {151-155},
          doi = {10.1038/s41550-017-0321-z},
archivePrefix = {arXiv},
       eprint = {1711.10609},
 primaryClass = {astro-ph.IM},
       adsurl = {https://ui.adsabs.harvard.edu/abs/2018NatAs...2..151N}
}

@ARTICLE{JamalBloom2020,
       author = {{Jamal}, Sara and {Bloom}, Joshua S.},
        title = "{On Neural Architectures for Astronomical Time-series Classification with Application to Variable Stars}",
      journal = {\apjs},
         year = 2020,
        month = oct,
       volume = {250},
       number = {2},
          eid = {30},
        pages = {30},
          doi = {10.3847/1538-4365/aba8ff},
archivePrefix = {arXiv},
       eprint = {2003.08618},
 primaryClass = {astro-ph.IM},
       adsurl = {https://ui.adsabs.harvard.edu/abs/2020ApJS..250...30J}
}

@ARTICLE{KimBailer-Jones2016,
       author = {{Kim}, Dae-Won and {Bailer-Jones}, Coryn A.~L.},
        title = "{A package for the automated classification of periodic variable stars}",
      journal = {\aap},
         year = 2016,
        month = mar,
       volume = {587},
          eid = {A18},
        pages = {A18},
          doi = {10.1051/0004-6361/201527188},
archivePrefix = {arXiv},
       eprint = {1512.01611},
 primaryClass = {astro-ph.IM},
       adsurl = {https://ui.adsabs.harvard.edu/abs/2016A&A...587A..18K}
}

@ARTICLE{Debosscher2007,
       author = {{Debosscher}, J. and {Sarro}, L.~M. and {Aerts}, C. and {Cuypers}, J. and {Vandenbussche}, B. and {Garrido}, R. and {Solano}, E.},
        title = "{Automated supervised classification of variable stars. I. Methodology}",
      journal = {\aap},
         year = 2007,
        month = dec,
       volume = {475},
       number = {3},
        pages = {1159-1183},
          doi = {10.1051/0004-6361:20077638},
archivePrefix = {arXiv},
       eprint = {0711.0703},
 primaryClass = {astro-ph},
       adsurl = {https://ui.adsabs.harvard.edu/abs/2007A&A...475.1159D}
}

@ARTICLE{Sarro2009,
       author = {{Sarro}, L.~M. and {Debosscher}, J. and {L{\'o}pez}, M. and {Aerts}, C.},
        title = "{Automated supervised classification of variable stars. II. Application to the OGLE database}",
      journal = {\aap},
         year = 2009,
        month = feb,
       volume = {494},
       number = {2},
        pages = {739-768},
          doi = {10.1051/0004-6361:200809918},
archivePrefix = {arXiv},
       eprint = {0806.3386},
 primaryClass = {astro-ph},
       adsurl = {https://ui.adsabs.harvard.edu/abs/2009A&A...494..739S}
}

@ARTICLE{Blomme2011,
       author = {{Blomme}, J. and {Sarro}, L.~M. and {O'Donovan}, F.~T. and {Debosscher}, J. and {Brown}, T. and {Lopez}, M. and {Dubath}, P. and {Rimoldini}, L. and {Charbonneau}, D. and {Dunham}, E. and {Mandushev}, G. and {Ciardi}, D.~R. and {De Ridder}, J. and {Aerts}, C.},
        title = "{Improved methodology for the automated classification of periodic variable stars}",
      journal = {\mnras},
         year = 2011,
        month = nov,
       volume = {418},
       number = {1},
        pages = {96-106},
          doi = {10.1111/j.1365-2966.2011.19466.x},
archivePrefix = {arXiv},
       eprint = {1101.5038},
 primaryClass = {astro-ph.IM},
       adsurl = {https://ui.adsabs.harvard.edu/abs/2011MNRAS.418...96B}
}

@ARTICLE{Muthukrishna2019,
       author = {{Muthukrishna}, Daniel and {Narayan}, Gautham and {Mandel}, Kaisey S. and {Biswas}, Rahul and {Hlo{\v{z}}ek}, Ren{\'e}e},
        title = "{RAPID: Early Classification of Explosive Transients Using Deep Learning}",
      journal = {\pasp},
         year = 2019,
        month = nov,
       volume = {131},
       number = {1005},
        pages = {118002},
          doi = {10.1088/1538-3873/ab1609},
archivePrefix = {arXiv},
       eprint = {1904.00014},
 primaryClass = {astro-ph.IM},
       adsurl = {https://ui.adsabs.harvard.edu/abs/2019PASP..131k8002M}
}

@ARTICLE{Pan2024,
       author = {{Pan}, Jia-Shu and {Ting}, Yuan-Sen and {Yu}, Jie},
        title = "{Astroconformer: The prospects of analysing stellar light curves with transformer-based deep learning models}",
      journal = {\mnras},
         year = 2024,
        month = mar,
       volume = {528},
       number = {4},
        pages = {5890-5903},
          doi = {10.1093/mnras/stae068},
archivePrefix = {arXiv},
       eprint = {2309.16316},
 primaryClass = {astro-ph.SR},
       adsurl = {https://ui.adsabs.harvard.edu/abs/2024MNRAS.528.5890P}
}

@misc{becker2025,
      title={Multiband Embeddings of Light Curves}, 
      author={I. Becker and P. Protopapas and M. Catelan and K. Pichara},
      year={2025},
      eprint={2501.12499},
      archivePrefix={arXiv},
      primaryClass={astro-ph.IM},
      url={https://arxiv.org/abs/2501.12499}, 
}

@ARTICLE{Richards2011,
       author = {{Richards}, Joseph W. and {Starr}, Dan L. and {Butler}, Nathaniel R. and {Bloom}, Joshua S. and {Brewer}, John M. and {Crellin-Quick}, Arien and {Higgins}, Justin and {Kennedy}, Rachel and {Rischard}, Maxime},
        title = "{On Machine-learned Classification of Variable Stars with Sparse and Noisy Time-series Data}",
      journal = {\apj},
         year = 2011,
        month = may,
       volume = {733},
       number = {1},
          eid = {10},
        pages = {10},
          doi = {10.1088/0004-637X/733/1/10},
archivePrefix = {arXiv},
       eprint = {1101.1959},
 primaryClass = {astro-ph.IM},
       adsurl = {https://ui.adsabs.harvard.edu/abs/2011ApJ...733...10R}
}

@ARTICLE{Vaswani2017,
       author = {{Vaswani}, Ashish and {Shazeer}, Noam and {Parmar}, Niki and {Uszkoreit}, Jakob and {Jones}, Llion and {Gomez}, Aidan N. and {Kaiser}, Lukasz and {Polosukhin}, Illia},
        title = "{Attention Is All You Need}",
      journal = {arXiv e-prints},
         year = 2017,
        month = jun,
          eid = {arXiv:1706.03762},
        pages = {arXiv:1706.03762},
          doi = {10.48550/arXiv.1706.03762},
archivePrefix = {arXiv},
       eprint = {1706.03762},
 primaryClass = {cs.CL},
       adsurl = {https://ui.adsabs.harvard.edu/abs/2017arXiv170603762V}
}

@ARTICLE{BERT,
       author = {{Devlin}, Jacob and {Chang}, Ming-Wei and {Lee}, Kenton and {Toutanova}, Kristina},
        title = "{BERT: Pre-training of Deep Bidirectional Transformers for Language Understanding}",
      journal = {arXiv e-prints},
         year = 2018,
        month = oct,
          eid = {arXiv:1810.04805},
        pages = {arXiv:1810.04805},
          doi = {10.48550/arXiv.1810.04805},
archivePrefix = {arXiv},
       eprint = {1810.04805},
 primaryClass = {cs.CL},
       adsurl = {https://ui.adsabs.harvard.edu/abs/2018arXiv181004805D}
}

@article{GPT,
  author = {Radford, Alec and Narasimhan, Karthik and Salimans, Tim and Sutskever, Ilya},
  title = {Improving language understanding by generative pre-training},
  year = 2018
}

@ARTICLE{Qingsong,
       author = {{Wen}, Qingsong and {Zhou}, Tian and {Zhang}, Chaoli and {Chen}, Weiqi and {Ma}, Ziqing and {Yan}, Junchi and {Sun}, Liang},
        title = "{Transformers in Time Series: A Survey}",
      journal = {arXiv e-prints},
         year = 2022,
        month = feb,
          eid = {arXiv:2202.07125},
        pages = {arXiv:2202.07125},
          doi = {10.48550/arXiv.2202.07125},
archivePrefix = {arXiv},
       eprint = {2202.07125},
 primaryClass = {cs.LG},
       adsurl = {https://ui.adsabs.harvard.edu/abs/2022arXiv220207125W}
}

@ARTICLE{GLU,
       author = {{Dauphin}, Yann N. and {Fan}, Angela and {Auli}, Michael and {Grangier}, David},
        title = "{Language Modeling with Gated Convolutional Networks}",
      journal = {arXiv e-prints},
         year = 2016,
        month = dec,
          eid = {arXiv:1612.08083},
        pages = {arXiv:1612.08083},
          doi = {10.48550/arXiv.1612.08083},
archivePrefix = {arXiv},
       eprint = {1612.08083},
 primaryClass = {cs.CL},
       adsurl = {https://ui.adsabs.harvard.edu/abs/2016arXiv161208083D}
}

@ARTICLE{MCD,
       author = {{Gal}, Yarin and {Ghahramani}, Zoubin},
        title = "{Dropout as a Bayesian Approximation: Representing Model Uncertainty in Deep Learning}",
      journal = {arXiv e-prints},
         year = 2015,
        month = jun,
          eid = {arXiv:1506.02142},
        pages = {arXiv:1506.02142},
          doi = {10.48550/arXiv.1506.02142},
archivePrefix = {arXiv},
       eprint = {1506.02142},
 primaryClass = {stat.ML},
       adsurl = {https://ui.adsabs.harvard.edu/abs/2015arXiv150602142G}
}

@ARTICLE{Temperature,
       author = {{Shen}, Jiajiang and {Wu}, Weiyan and {Xu}, Qianyu},
        title = "{Accurate Prediction of Temperature Indicators in Eastern China Using a Multi-Scale CNN-LSTM-Attention model}",
      journal = {arXiv e-prints},
         year = 2024,
        month = dec,
          eid = {arXiv:2412.07997},
        pages = {arXiv:2412.07997},
          doi = {10.48550/arXiv.2412.07997},
archivePrefix = {arXiv},
       eprint = {2412.07997},
 primaryClass = {cs.LG},
       adsurl = {https://ui.adsabs.harvard.edu/abs/2024arXiv241207997S}
}

@ARTICLE{Gasoline,
       author = {{Ranjbar}, Mahmoud and {Rahimzadeh}, Mohammad},
        title = "{Advancing Gasoline Consumption Forecasting: A Novel Hybrid Model Integrating Transformers, LSTM, and CNN}",
      journal = {arXiv e-prints},
         year = 2024,
        month = oct,
          eid = {arXiv:2410.16336},
        pages = {arXiv:2410.16336},
          doi = {10.48550/arXiv.2410.16336},
archivePrefix = {arXiv},
       eprint = {2410.16336},
 primaryClass = {eess.SY},
       adsurl = {https://ui.adsabs.harvard.edu/abs/2024arXiv241016336R}
}

@article{Stocks,
	article-number = {1985},
	author = {Zhang, Jilin and Ye, Lishi and Lai, Yongzeng},
	doi = {10.3390/math11091985},
	issn = {2227-7390},
	journal = {Mathematics},
	number = {9},
	title = {Stock Price Prediction Using CNN-BiLSTM-Attention Model},
	url = {https://www.mdpi.com/2227-7390/11/9/1985},
	volume = {11},
	year = {2023}}

@article{Hochreiter1997,
    author = {Hochreiter, Sepp and Schmidhuber, Jürgen},
    title = {Long Short-Term Memory},
    journal = {Neural Computation},
    volume = {9},
    number = {8},
    pages = {1735-1780},
    year = {1997},
    month = {11},
    issn = {0899-7667},
    doi = {10.1162/neco.1997.9.8.1735},
    url = {https://doi.org/10.1162/neco.1997.9.8.1735},
    eprint = {https://direct.mit.edu/neco/article-pdf/9/8/1735/813796/neco.1997.9.8.1735.pdf},
}

@ARTICLE{Schuster1997,
  author={Schuster, M. and Paliwal, K.K.},
  journal={IEEE Transactions on Signal Processing}, 
  title={Bidirectional recurrent neural networks}, 
  year={1997},
  volume={45},
  number={11},
  pages={2673-2681},
  doi={10.1109/78.650093}}

@article{Srivastava2014,
author = {Srivastava, Nitish and Hinton, Geoffrey and Krizhevsky, Alex and Sutskever, Ilya and Salakhutdinov, Ruslan},
title = {Dropout: a simple way to prevent neural networks from overfitting},
year = {2014},
issue_date = {January 2014},
publisher = {JMLR.org},
volume = {15},
number = {1},
issn = {1532-4435},
journal = {J. Mach. Learn. Res.},
month = jan,
pages = {1929–1958},
numpages = {30}
}

@ARTICLE{2018AJ....156..123A,
       author = {{Astropy Collaboration} and {Price-Whelan}, A.~M. and {Sip{\H{o}}cz}, B.~M. and {G{\"u}nther}, H.~M. and {Lim}, P.~L. and {Crawford}, S.~M. and {Conseil}, S. and {Shupe}, D.~L. and {Craig}, M.~W. and {Dencheva}, N. and {Ginsburg}, A. and {VanderPlas}, J.~T. and {Bradley}, L.~D. and {P{\'e}rez-Su{\'a}rez}, D. and {de Val-Borro}, M. and {Aldcroft}, T.~L. and {Cruz}, K.~L. and {Robitaille}, T.~P. and {Tollerud}, E.~J. and {Ardelean}, C. and {Babej}, T. and {Bach}, Y.~P. and {Bachetti}, M. and {Bakanov}, A.~V. and {Bamford}, S.~P. and {Barentsen}, G. and {Barmby}, P. and {Baumbach}, A. and {Berry}, K.~L. and {Biscani}, F. and {Boquien}, M. and {Bostroem}, K.~A. and {Bouma}, L.~G. and {Brammer}, G.~B. and {Bray}, E.~M. and {Breytenbach}, H. and {Buddelmeijer}, H. and {Burke}, D.~J. and {Calderone}, G. and {Cano Rodr{\'\i}guez}, J.~L. and {Cara}, M. and {Cardoso}, J.~V.~M. and {Cheedella}, S. and {Copin}, Y. and {Corrales}, L. and {Crichton}, D. and {D'Avella}, D. and {Deil}, C. and {Depagne}, {\'E}. and {Dietrich}, J.~P. and {Donath}, A. and {Droettboom}, M. and {Earl}, N. and {Erben}, T. and {Fabbro}, S. and {Ferreira}, L.~A. and {Finethy}, T. and {Fox}, R.~T. and {Garrison}, L.~H. and {Gibbons}, S.~L.~J. and {Goldstein}, D.~A. and {Gommers}, R. and {Greco}, J.~P. and {Greenfield}, P. and {Groener}, A.~M. and {Grollier}, F. and {Hagen}, A. and {Hirst}, P. and {Homeier}, D. and {Horton}, A.~J. and {Hosseinzadeh}, G. and {Hu}, L. and {Hunkeler}, J.~S. and {Ivezi{\'c}}, {\v{Z}}. and {Jain}, A. and {Jenness}, T. and {Kanarek}, G. and {Kendrew}, S. and {Kern}, N.~S. and {Kerzendorf}, W.~E. and {Khvalko}, A. and {King}, J. and {Kirkby}, D. and {Kulkarni}, A.~M. and {Kumar}, A. and {Lee}, A. and {Lenz}, D. and {Littlefair}, S.~P. and {Ma}, Z. and {Macleod}, D.~M. and {Mastropietro}, M. and {McCully}, C. and {Montagnac}, S. and {Morris}, B.~M. and {Mueller}, M. and {Mumford}, S.~J. and {Muna}, D. and {Murphy}, N.~A. and {Nelson}, S. and {Nguyen}, G.~H. and {Ninan}, J.~P. and {N{\"o}the}, M. and {Ogaz}, S. and {Oh}, S. and {Parejko}, J.~K. and {Parley}, N. and {Pascual}, S. and {Patil}, R. and {Patil}, A.~A. and {Plunkett}, A.~L. and {Prochaska}, J.~X. and {Rastogi}, T. and {Reddy Janga}, V. and {Sabater}, J. and {Sakurikar}, P. and {Seifert}, M. and {Sherbert}, L.~E. and {Sherwood-Taylor}, H. and {Shih}, A.~Y. and {Sick}, J. and {Silbiger}, M.~T. and {Singanamalla}, S. and {Singer}, L.~P. and {Sladen}, P.~H. and {Sooley}, K.~A. and {Sornarajah}, S. and {Streicher}, O. and {Teuben}, P. and {Thomas}, S.~W. and {Tremblay}, G.~R. and {Turner}, J.~E.~H. and {Terr{\'o}n}, V. and {van Kerkwijk}, M.~H. and {de la Vega}, A. and {Watkins}, L.~L. and {Weaver}, B.~A. and {Whitmore}, J.~B. and {Woillez}, J. and {Zabalza}, V. and {Astropy Contributors}},
        title = "{The Astropy Project: Building an Open-science Project and Status of the v2.0 Core Package}",
      journal = {\aj},
         year = 2018,
        month = sep,
       volume = {156},
       number = {3},
          eid = {123},
        pages = {123},
          doi = {10.3847/1538-3881/aabc4f},
archivePrefix = {arXiv},
       eprint = {1801.02634},
 primaryClass = {astro-ph.IM},
       adsurl = {https://ui.adsabs.harvard.edu/abs/2018AJ....156..123A}
}

@ARTICLE{2013A&A...558A..33A,
       author = {{Astropy Collaboration} and {Robitaille}, Thomas P. and
         {Tollerud}, Erik J. and {Greenfield}, Perry and {Droettboom}, Michael and
         {Bray}, Erik and {Aldcroft}, Tom and {Davis}, Matt and
         {Ginsburg}, Adam and {Price-Whelan}, Adrian M. and
         {Kerzendorf}, Wolfgang E. and {Conley}, Alexander and {Crighton}, Neil and
         {Barbary}, Kyle and {Muna}, Demitri and {Ferguson}, Henry and
         {Grollier}, Fr{\'e}d{\'e}ric and {Parikh}, Madhura M. and
         {Nair}, Prasanth H. and {Unther}, Hans M. and {Deil}, Christoph and
         {Woillez}, Julien and {Conseil}, Simon and {Kramer}, Roban and
         {Turner}, James E.~H. and {Singer}, Leo and {Fox}, Ryan and
         {Weaver}, Benjamin A. and {Zabalza}, Victor and {Edwards}, Zachary I. and
         {Azalee Bostroem}, K. and {Burke}, D.~J. and {Casey}, Andrew R. and
         {Crawford}, Steven M. and {Dencheva}, Nadia and {Ely}, Justin and
         {Jenness}, Tim and {Labrie}, Kathleen and {Lim}, Pey Lian and
         {Pierfederici}, Francesco and {Pontzen}, Andrew and {Ptak}, Andy and
         {Refsdal}, Brian and {Servillat}, Mathieu and {Streicher}, Ole},
        title = "{Astropy: A community Python package for astronomy}",
      journal = {\aap},
         year = "2013",
        month = "Oct",
       volume = {558},
          eid = {A33},
        pages = {A33},
          doi = {10.1051/0004-6361/201322068},
archivePrefix = {arXiv},
       eprint = {1307.6212},
 primaryClass = {astro-ph.IM},
       adsurl = {https://ui.adsabs.harvard.edu/abs/2013A&A...558A..33A}
}

@software{lightkurve,
       author = {{Lightkurve Collaboration} and {Cardoso}, Jos{\'e} Vin{\'\i}cius de Miranda and {Hedges}, Christina and {Gully-Santiago}, Michael and {Saunders}, Nicholas and {Cody}, Ann Marie and {Barclay}, Thomas and {Hall}, Oliver and {Sagear}, Sheila and {Turtelboom}, Emma and {Zhang}, Johnny and {Tzanidakis}, Andy and {Mighell}, Ken and {Coughlin}, Jeff and {Bell}, Keaton and {Berta-Thompson}, Zach and {Williams}, Peter and {Dotson}, Jessie and {Barentsen}, Geert},
        title = "{Lightkurve: Kepler and TESS time series analysis in Python}",
 howpublished = {Astrophysics Source Code Library, record ascl:1812.013},
         year = 2018,
        month = dec,
          eid = {ascl:1812.013},
       adsurl = {https://ui.adsabs.harvard.edu/abs/2018ascl.soft12013L}
}

@ARTICLE{Roxburgh2025,
       author = {{Roxburgh}, Hugh and {Ridden-Harper}, Ryan and {Moore}, Andrew and {Montilla}, Clarinda and {Leicester}, Brayden and {Lane}, Zachary G. and {Freeburn}, James and {Rest}, Armin and {Bannister}, Michele T. and {Ridden-Harper}, Andrew R. and {Hubley}, Lancia and {Wang}, Qinan and {Hounsell}, Rebekah and {Cooke}, Jeff and {Coulter}, Dave A. and {Fausnaugh}, Michael M.},
        title = "{TESSELLATE: Piecing Together the Variable Sky With TESS}",
      journal = {arXiv e-prints},
         year = 2025,
        month = feb,
          eid = {arXiv:2502.16905},
        pages = {arXiv:2502.16905},
archivePrefix = {arXiv},
       eprint = {2502.16905},
 primaryClass = {astro-ph.IM},
       adsurl = {https://ui.adsabs.harvard.edu/abs/2025arXiv250216905R}
}

@ARTICLE{Olmschenk2024,
       author = {{Olmschenk}, Greg and {Barry}, Richard K. and {Ishitani Silva}, Stela and {Schnittman}, Jeremy D. and {Cieplak}, Agnieszka M. and {Powell}, Brian P. and {Kruse}, Ethan and {Barclay}, Thomas and {Solanki}, Siddhant and {Ortega}, Bianca and {Baker}, John and {Yesenia Helem Salinas}, Mamani},
        title = "{Short-period Variables in TESS Full-frame Image Light Curves Identified via Convolutional Neural Networks}",
      journal = {\aj},
         year = 2024,
        month = aug,
       volume = {168},
       number = {2},
          eid = {83},
        pages = {83},
          doi = {10.3847/1538-3881/ad55f1},
archivePrefix = {arXiv},
       eprint = {2402.12369},
 primaryClass = {astro-ph.SR},
       adsurl = {https://ui.adsabs.harvard.edu/abs/2024AJ....168...83O}
}

@ARTICLE{lomb1976,
   author = {{Lomb}, N.~R.},
    title = "{Least-squares frequency analysis of unequally spaced data}",
  journal = {\apss},
     year = 1976,
    month = feb,
   volume = 39,
    pages = {447-462},
      doi = {10.1007/BF00648343},
   adsurl = {http://adsabs.harvard.edu/abs/1976Ap%26SS..39..447L}
}

@ARTICLE{scargle1982,
   author = {{Scargle}, J.~D.},
    title = "{Studies in astronomical time series analysis. II - Statistical aspects of spectral analysis of unevenly spaced data}",
  journal = {\apj},
     year = 1982,
    month = dec,
   volume = 263,
    pages = {835-853},
      doi = {10.1086/160554},
   adsurl = {http://adsabs.harvard.edu/abs/1982ApJ...263..835S}
}

@article{Shannon1948,
    author = {Shannon, C. E.},
    title = {A Mathematical Theory of Communication},
    journal = {Bell System Technical Journal},
    volume = {27},
    number = {4},
    pages = {623-656},
    doi = {10.1002/j.1538-7305.1948.tb00917.x},
    url = {https://onlinelibrary.wiley.com/doi/abs/10.1002/j.1538-7305.1948.tb00917.x},
    eprint = {https://onlinelibrary.wiley.com/doi/pdf/10.1002/j.1538-7305.1948.tb00917.x},
    year = {1948}
}

@article{friedman2001,
  title={Greedy function approximation: a gradient boosting machine},
  author={Friedman, Jerome H},
  journal={Annals of statistics},
  pages={1189--1232},
  year={2001},
  publisher={JSTOR}
}

@ARTICLE{Breiman:fb,
author = {Breiman, Leo},
title = {{Random Forests}},
journal = {Machine Learning},
year = {2001},
volume = {45},
number = {1},
pages = {5--32},
month = oct
}

@ARTICLE{Aerts2021,
       author = {{Aerts}, C.},
        title = "{Probing the interior physics of stars through asteroseismology}",
      journal = {Reviews of Modern Physics},
         year = 2021,
        month = jan,
       volume = {93},
       number = {1},
          eid = {015001},
        pages = {015001},
          doi = {10.1103/RevModPhys.93.015001},
archivePrefix = {arXiv},
       eprint = {1912.12300},
 primaryClass = {astro-ph.SR},
       adsurl = {https://ui.adsabs.harvard.edu/abs/2021RvMP...93a5001A}
}

@BOOK{asteroseismology,
       author = {{Aerts}, Conny and {Christensen-Dalsgaard}, J{\o}rgen and {Kurtz}, Donald W.},
        title = "{Asteroseismology}",
         year = 2010,
          doi = {10.1007/978-1-4020-5803-5},
       adsurl = {https://ui.adsabs.harvard.edu/abs/2010aste.book.....A}
}

@ARTICLE{Kurtz2022,
       author = {{Kurtz}, Donald W.},
        title = "{Asteroseismology Across the Hertzsprung-Russell Diagram}",
      journal = {\araa},
         year = 2022,
        month = aug,
       volume = {60},
        pages = {31-71},
          doi = {10.1146/annurev-astro-052920-094232},
       adsurl = {https://ui.adsabs.harvard.edu/abs/2022ARA&A..60...31K}
}

@ARTICLE{Nascimbeni2025,
       author = {{Nascimbeni}, V. and {Piotto}, G. and {Cabrera}, J. and {Montalto}, M. and {Marinoni}, S. and {Marrese}, P.~M. and {Aerts}, C. and {Altavilla}, G. and {Benatti}, S. and {B{\"o}rner}, A. and {Deleuil}, M. and {Desidera}, S. and {Gizon}, L. and {Goupil}, M.~J. and {Granata}, V. and {Heras}, A.~M. and {Magrin}, D. and {Malavolta}, L. and {Mas-Hesse}, J.~M. and {Osborn}, H.~P. and {Pagano}, I. and {Paproth}, C. and {Pollacco}, D. and {Prisinzano}, L. and {Ragazzoni}, R. and {Ramsay}, G. and {Rauer}, H. and {Tkachenko}, A. and {Udry}, S.},
        title = "{The PLATO field selection process: II. Characterization of LOPS2, the first long-pointing field}",
      journal = {\aap},
         year = 2025,
        month = feb,
       volume = {694},
          eid = {A313},
        pages = {A313},
          doi = {10.1051/0004-6361/202452325},
archivePrefix = {arXiv},
       eprint = {2501.07687},
 primaryClass = {astro-ph.EP},
       adsurl = {https://ui.adsabs.harvard.edu/abs/2025A&A...694A.313N}
}

@ARTICLE{Jannsen2025,
       author = {{Jannsen}, N. and {Tkachenko}, A. and {Royer}, P. and {De Ridder}, J. and {Seynaeve}, D. and {Aerts}, C. and {Aigrain}, S. and {Plachy}, E. and {Bodi}, A. and {Uzundag}, M. and {Bowman}, D.~M. and {Fritzewski}, D.~J. and {IJspeert}, L.~W. and {Li}, G. and {Pedersen}, M.~G. and {Vanrespaille}, M. and {Van Reeth}, T.},
        title = "{MOCKA {\textendash} A PLATO mock asteroseismic catalogue: Simulations for gravity-mode oscillators}",
      journal = {\aap},
         year = 2025,
        month = feb,
       volume = {694},
          eid = {A185},
        pages = {A185},
          doi = {10.1051/0004-6361/202452811},
archivePrefix = {arXiv},
       eprint = {2412.10508},
 primaryClass = {astro-ph.SR},
       adsurl = {https://ui.adsabs.harvard.edu/abs/2025A&A...694A.185J}
}

@ARTICLE{Audenaert2025,
       author = {{Audenaert}, Jeroen},
        title = "{From stellar light to astrophysical insight: automating variable star research with machine learning}",
      journal = {\apss},
         year = 2025,
        month = jul,
       volume = {370},
       number = {7},
          eid = {72},
        pages = {72},
          doi = {10.1007/s10509-025-04460-5},
archivePrefix = {arXiv},
       eprint = {2507.03093},
 primaryClass = {astro-ph.IM},
       adsurl = {https://ui.adsabs.harvard.edu/abs/2025Ap&SS.370...72A}
}

@ARTICLE{QLP2,
       author = {{Huang}, Chelsea X. and {Vanderburg}, Andrew and {P{\'a}l}, Andras and {Sha}, Lizhou and {Yu}, Liang and {Fong}, Willie and {Fausnaugh}, Michael and {Shporer}, Avi and {Guerrero}, Natalia and {Vanderspek}, Roland and {Ricker}, George},
        title = "{Photometry of 10 Million Stars from the First Two Years of TESS Full Frame Images: Part II}",
      journal = {Research Notes of the American Astronomical Society},
         year = 2020,
        month = nov,
       volume = {4},
       number = {11},
          eid = {206},
        pages = {206},
          doi = {10.3847/2515-5172/abca2d},
       adsurl = {https://ui.adsabs.harvard.edu/abs/2020RNAAS...4..206H}
}
\bibliographystyle{aasjournal}



\end{document}